\documentstyle[amssymb,stmaryrd,epsfig]{mnv2}
   
\title[Isolated Galaxies]
{The Photometric Properties of Isolated Early-Type Galaxies}

\author[Reda, Forbes, Beasley, O'Sullivan \& Goudfrooij]{
Fatma M. Reda$^{1,2}$\thanks{freda@astro.swin.edu.au}
Duncan A. Forbes$^{1}$\thanks{dforbes@swin.edu.au}
Michael A. Beasley$^{1,3}$\thanks{mbeasley@ucolick.org} \\
\LARGE
Ewan J. O'Sullivan$^{4}$\thanks{ejos@head-cfa.cfa.harvard.edu} 
Paul Goudfrooij$^{5}$\thanks{goudfroo@stsci.edu}\\
  $^1$Centre for Astrophysics \& Supercomputing, Swinburne University, 
Hawthorn, VIC 3122, Australia\\
  $^2$Astronomy Department, National Research Institute of Astronomy and 
Geophysics, Helwan, Cairo 11421, Egypt\\
  $^3$Lick Observatory, University of California, Santa Cruz, 
CA 95064, USA\\
  $^4$Harvard-Smithsonian Center for Astrophysics, MS-67, 60
Garden Street, Cambridge, MA 02138, USA\\
  $^5$Space Telescope Science Institute, 3700 San Martin Drive,
Baltimore, MD 21218, USA\\
}
\begin{document} 
\date{Accepted 2004 July 21. Received 2004 June 19}

\pagerange{\pageref{firstpage}--\pageref{lastpage}} \pubyear{2004}
\maketitle  
  
\begin{abstract}
Isolated galaxies are important since they probe the lowest density regimes 
inhabited by galaxies.
We define a sample of 36 nearby isolated early-type galaxies for
further study. Our isolation criteria require them to have no
comparable-mass neighbours within 2 $B$-band magnitudes, 
0.67 Mpc in the plane of the sky and 700 km s$^{-1}$ in recession velocity.
New wide-field optical imaging of 10 isolated galaxies with the
Anglo-Australian Telescope confirms their
early-type morphology and relative isolation. We also present
imaging of 4 galaxy groups as a control sample. The isolated
galaxies are shown to be more gravitationally isolated than the
group galaxies. 
We find that the isolated early-type galaxies have a mean effective
colour of ($B$--$R$)$_e = 1.54 \pm 0.14$, similar to their high-density
counterparts. They reveal a similar colour-magnitude relation
slope and small intrinsic scatter to cluster ellipticals. They
also follow the Kormendy relation of surface brightness versus
size for luminous cluster galaxies. Such properties suggest that
the isolated galaxies formed at a similar epoch to cluster
galaxies, such that the bulk of their stars are very old. However,
our galaxy modelling reveals evidence for dust lanes, plumes,
shells, boxy and disk isophotes in four out of nine galaxies. 
Thus at least some isolated galaxies have experienced a recent 
merger/accretion event which may have induced a small 
burst of star formation. 
We derive luminosity functions for the isolated galaxies and
find a faint slope of --1.2, which is similar to the `universal'
slope found in a wide variety of environments. 
We examine the number density distribution of galaxies in the field of the
isolated galaxies. Only the very faintest dwarf galaxies (M$_R \ga -15.5$)
appear to be associated with the isolated galaxies, whereas any
intermediate-luminosity galaxies appear to lie in the
background. 
Finally we discuss possible formation scenarios for isolated
early-type galaxies. Early epoch formation and a merger/accretion of galaxies 
are possible explanations.
The collapse of a large, virialised group is an
unlikely explanation, but that of a poor group remains viable. 

\end{abstract}
\begin{keywords}  
galaxies: elliptical and lenticular - galaxies: photometry - galaxies: 
structure - galaxies: luminosity function, mass function.
\end{keywords}


\section{Introduction}

Early-type galaxies are more commonly found in dense environments such as 
clusters and groups, than in the field. Within a cluster, the dense core 
regions are richer with early-type galaxies than the lower-density outskirts. 
This distribution of galaxy morphology with environment is known as the 
morphology-density relation (Dressler 1980). In all environments, early-type 
galaxies 
show morphological fine structures such as tidal tails, shells, dust lanes, 
disky or boxy components, etc. (e.g.\ Schweizer et al. 1990). Theoretical 
models often attempt to explain 
these morphological structures as a result of past merging in earlier 
galaxy formation stages or, alternatively, of tidal interactions 
with close neighbours. 
In dense cluster environments, merging was more likely at earlier epochs 
when the cluster was still forming and velocity dispersions were low enough 
to allow outright mergers. 

Studies of the stellar populations in early-type galaxies in different 
environments suggest that field 
galaxies tend to be bluer, and have spectral indices indicating slightly 
younger mean stellar ages than their cluster counterparts (e.g. De Carvalho \& 
Djorgovski 1992; Terlevich \& Forbes 2002; Girardi et al. 2003).
For a sample of 35 field early-type galaxies at a median redshift of 0.9, 
Kodama, Bower \& Bell (1999) found some of them to be redder in their 
colour-magnitude relation (CMR) than that for Coma cluster galaxies. 
These authors concluded
that at least half of the field early-type galaxies at these redshifts 
formed their stars at $z>2$.
The Kodama et al. sample of high redshift field galaxies shows a greater
scatter in the CMR than that obtained by Stanford, 
Eisenhardt \&  Dickinson (1998) for a cluster at 
$z\sim0.9$. This suggests a slightly more extended star formation period for 
early-type galaxies in the field than in clusters. 

Isolated galaxies represent the `extreme field' galaxy population and thus 
provide additional insight to those environment related processes. 
A carefully selected sample of isolated early-type galaxies will offer a 
useful tool to tackle many of the outstanding issues in galaxy formation. 
For example, in dense environments, separating the effects of 
secular and environmental evolution (``nature versus nurture'') is challenging.
However, in such extremely low-density environments as those
of isolated galaxies, one can eliminate the interaction 
processes that affect the evolution of galaxies in high-density environments
such as ram-pressure stripping (Gunn \& Gott 1972), strangulation (Balogh  \& 
Morris 2000; Fujita 2004), high-speed 
galaxy-galaxy encounters and tidal interactions with the cluster gravitational 
potential (Moore et al. 1996, 1999; Moore, Lake \& Katz 1998).
Previous studies such as Karachentseva (1973), Colbert, Mulchaey \& Zabludoff 
(2001), Aars, Marcum \& Fanelli (2001), Kuntschner et al. (2002), 
Smith \& Martinez (2003) and Stocke et al. (2004) have introduced 
samples of early-type galaxies that are in 
low-density environments or are isolated from neighbours. 
Each study has its own advantages and limitations. 
In this paper, we define a new sample of 36 candidate isolated early-type 
galaxies with strict isolation conditions to ensure the absence of any strong 
tidal effects from a massive neighbour. 
Using the Wide Field Imager (WFI) on the 3.9-m Anglo-Australian Telescope 
(AAT) we obtained imaging in the $B$ and $R$ filters for a wide area 
surrounding 
10 galaxies in the sample. This allows us to study the galaxy surface 
density distribution and photometric parameters of the faint galaxies 
in these fields.  

In section 2, we give a summary of the previous 
studies of isolated early-type galaxies stating their selection 
criterion and briefly mention their results. Our selection 
criteria and the sample candidates are presented in section 3. 
Observations and data reduction for a subsample of 10 isolated galaxies and 
a comparison sample, are discussed in section 4, while their photometric 
measurements as well as modelling, residual images and isophotal shape 
parameters are discussed in section 5. 
The detections of galaxies in the field  
are presented in section 6. Finally, we discuss the results and possible 
formation scenarios for isolated early-type galaxies in section 7.
Throughout this paper we assume a Hubble constant of 
H$_0$ = 75 km s$^{-1}$ Mpc$^{-1}$.

\section{Previous studies of isolated galaxies}

Here we briefly review previous studies of galaxies in low density 
environments. Each study had its own advantages and limitations.

{\it Karachentseva (1973)} - a catalogue of $\sim$1000 galaxies with 
 m$_B<15.7$ and declination $>-3^\circ$. The catalogue was chosen from the 
Zwicky et al. (1957) catalogue which is known to be biased against 
low surface brightness galaxies. 
The environment of 
each galaxy was inspected using the Palomar Observatory Sky Survey (POSS) for 
large neighbours. Isolation was established as the absence of companions 
within 20 galaxy diameters that have less than a 
$\pm2$ magnitude difference from 
the primary galaxy (Stocke et al. 2004). Due to saturation, poor resolution or 
other reasons in the POSS plates, one half of ellipticals in this catalogue 
appear to be misclassified (Saucedo-Morales \& Bieging 2001).
   
{\it Reduzzi, Longhetti \& Rampazzo (1996)} - a sample of 61 galaxies was
selected from `The Surface Photometry Catalogue' of the ESO-Uppsala Galaxies 
survey (Lauberts \& Valentijn 1989). 
The galaxies were considered as isolated when the average number of galaxies 
of any morphology type was $\leq 1.5$ within a radius of $1^{\circ}$.
The sample of 61 galaxies were imaged for 20 min each at the 0.9-m ESO-Dutch 
telescope at La Silla, Chile.
They noted that ten of the galaxies (i.e., 16\% of their galaxy sample)
contained spiral arms or bars, indicating they were of late-type morphology.
 The isolation of the galaxies was quantified using the ratio of the
separation between the target and the nearest galaxy in units of the target 
galaxy diameter, i.e.\ S/$D_{25}$. 
For galaxies with S/$D_{25} \leq 20$, they examined an additional parameter, 
i.e. the luminosity ratio between the target and the nearest galaxy 
$L_B/L_B^n$. 
Using these two parameters S/$D_{25}$ and $L_B/L_B^n$, they suggested that 
three galaxies had significant companions. One of these was a  
 misclassified spiral. 
Based on the S/$D_{25}$ and $L_B/L_B^n$ values given in their paper, we 
determined 
that another 7 galaxies have significant companions. This leaves a total of 42 
galaxies that can be considered isolated and of early-type. 
For their total sample of 61  galaxies, they 
found about 40\% to have fine structures such as shells, dust, tidal tails, 
etc. Considering only the 42 isolated early-type galaxies, this 
percentage increases to 52\%.

{\it Colbert, Mulchaey \& Zabludoff (2001)} - a sample of 30 early-type 
galaxies, selected from the Third Reference Catalog 
(RC3; de Vaucouleurs et al. 1991) catalogue. 
Galaxies were selected to have no catalogued neighbours within 
1 h$_{100}^{-1}$ Mpc and $\pm$1000 km s$^{-1}$. Imaging
of fields surrounding each galaxy was then used to confirm the lack of
large galaxies within 200 h$_{100}^{-1}$ kpc. However, some of the target
galaxies themselves had luminosities only slightly greater than the 
catalogue limit. This means that some candidate isolated galaxies actually
have nearby neighbours of quite similar luminosity.
Our examination of this sample with the Digitized Sky Survey (DSS) suggests 
that $\sim$50\% of the candidate isolated galaxies have
nearby neighbours. 
       
{\it Aars, Marcum \&  Fanelli (2001)} -
a sample of nine isolated 
elliptical galaxies selected from Karachentseva (1973). 
Two selection criteria were adopted for the initial identification of the 
isolated elliptical candidates. Firstly, elliptical galaxies were selected 
from the NASA Extragalactic Database (NED) that have separations of 2.5 Mpc 
from any other galaxy in the RC3.
Secondly, the galaxy should have no neighbours with a known redshift (from
 NED) brighter than M$_V=-16.5$, within a projected separation of 2.5 Mpc.    
 Only one galaxy in the RC3 and 13 in the Karachentseva (1973)
 catalogue meet these isolation criteria. 
 From this sample of 14 galaxies, wide-field CCD
 images in the $V$-band were obtained using the 2.1-m telescope at
 McDonald Observatory. The one RC3  
galaxy was a southern 
object which could not be imaged from McDonald Observatory and was therefore 
not included in the final sample. The images were used to check the isolation 
of these galaxies and to confirm the morphological type. This process 
identified that three were actually spiral galaxies and one was an 
irregular-type galaxy. These were excluded from further study leaving a final 
sample of nine isolated elliptical galaxies. 
Aars et al. defined a characteristic number density of projected galaxies 
on the sky for known loose groups and clusters. 
Comparing the number density of galaxies detected in the 
field of their 9 galaxies with these characteristic densities, they 
identified 5 galaxies to be in environments similar to those 
of loose groups. The environments of the remaining 
four galaxies were confirmed to be of low density.
    
{\it Kuntschner et al. (2002)} -a sample of nine nearby early-type 
galaxies in low-density environments (strictly speaking this is not an 
isolated galaxy sample). They were selected from the 
Hydra-Centaurus Catalogue of Raychaudhury (1989, 1990) to be 
early-type (T $<-3$) with velocities $<7,000$ km s$^{-1}$ and apparent 
magnitudes of b$_J\leq 16.1$. Their sample completeness at this magnitude is 
$\sim 50\%$. 
These galaxies were then required to have a maximum of two neighbours 
with b$_J\leq 16.7$, 
within a radius of 1.3 Mpc and a velocity difference 350 km s$^{-1}$. 
The result was a sample of 40 E and S0 galaxies. 
Visual inspection of the DSS images revealed that some of the galaxies have 
late-type morphologies leaving a sample of 30 early-type galaxies 
in low-density environments. Spectroscopic observations of 24 galaxies of this 
sample were obtained at the 2.3-m telescope at the Siding Spring 
Observatory, Australia. Kuntschner et al. found 5 galaxies to have
emission line spectra typical of spiral galaxies, 3 have red shifts beyond 
7,000 km s$^{-1}$ (the adopted redshift limit in their work), 4 galaxies 
were classified as group/cluster members 
and 2 had spectra of poor signal-to-noise.
Kuntschner et al. also obtained optical $U$$V$$R$ and near-infrared $K$$_s$ 
imaging data for their sample with the CTIO 1.5-m telescope in Chile. 
By inspecting the 
model-subtracted images of the sample, one galaxy showed clearly visible 
spiral arms. In their final sample of 9 galaxies, 6 are members of the 
Arp $\&$ Madore (1987) catalogue of peculiar galaxies, 
i.e. they have indications of a past merger. 
The spectra indicate the presence of young stellar populations in 
several of these galaxies.

{\it Smith, Martinez \& Graham (2003)} - a sample of 32 isolated early-type 
galaxies (T $<-4$). The Lyon-Meudon Extragalactic Data Archive (LEDA) 
catalogue was used to find ellipticals which have 
velocities $<10,000$ km s$^{-1}$, absolute magnitudes
M$_B$$\leq -19$ and Galactic latitudes $b>\mid 25^\circ\mid$. The LEDA 
 database was also used to identify the faint neighbours in the field of the 
selected galaxies. The isolation criteria include a $B$-band magnitude 
difference $>0.7$ between the primary galaxy and the neighbouring galaxies 
 within a projected distance of 1 Mpc, or $> 2.2$ magnitudes within 500
 kpc. They did not include any redshift information in their isolation
 criteria. They used the published data from the  UK and Palomar
 Schmidt Sky Survey plates to select all detected dwarf galaxies with
 absolute magnitudes of M$_{B}\leq -14.6$ in the field of the  
primary galaxies. The sample of dwarfs brighter than M$_B=-16.8$ or with 
low surface brightness are incomplete.

{\it Stocke et al. (2004)} - a sample of early-type galaxies were 
selected from the Karachentseva (1973) catalogue. Thirteen galaxies were 
eliminated from the initially selected sample after checking the POSS 
plates and finding some comparably sized companions missed by 
Karachentseva (1973).
Some of the sample galaxies are too faint and compact to be classified 
correctly by POSS. So for $80\%$ of the Es and 86\% of the S0s galaxies, the 
morphology and isolation were examined by Stocke et al. using optical imaging 
taken at Mt. 
Hopkins 0.6-m telescope and from images included in Adams, Jensen \& 
Stocke (1980). 
That lead to the elimination of a further seven galaxies. Thus the final 
sample consists of 62 E and 36 S0 galaxies.

Having decided that some of the previous isolated galaxy samples 
were not suitable for our purpose, we have defined a new sample.

\begin{table*}
 \begin{minipage}{180mm}
\begin{center}
\begin{tabular}{llclcl}
\multicolumn{6}{l}{\bf Table 1. \small The sample of 36 isolated galaxies.}\\
\hline
 Galaxy & Type & $B$$_T$ & Magnitude & Dist. & Previous  \\
        &      & (mag) & source  & (Mpc) & samples \\
\hline
NGC  682       & E/S0   &   14.36 & LEDA  &   73  &   \\
NGC  821       & E      &   11.33 & LEDA  &   23  & SMG03  \\
NGC 1041       & E/S0   &   14.28 & LEDA  &   93  &   \\
NGC 1045       & E/S0   &   13.45 & LEDA  &   60  &   \\ 
NGC 1132       & E      &   13.03 & LEDA  &   92  & CMZ01  \\
NGC 1162       & E      &   12.88 & LEDA  &   51  & SMG03  \\ 
NGC 2110       & E/S0   &   12.21 & LEDA  &   28  & CMZ01  \\  
NGC 2128       & E/S0   &   12.66 & NED   &   44  &   \\    
NGC 2271       & E/S0   &   12.52 & LEDA  &   32  &   \\   
NGC 2865       & E      &   11.98 & LEDA  &   35  & RLR96  \\    
NGC 3562       & E      &   12.99 & LEDA  &   93  &   \\   
NGC 4240       & E      &   13.31 & LEDA  &   26  &   \\    
NGC 4271       & E/S0   &   13.48 & NED   &   66  &   \\    
NGC 4555       & E      &   13.05 & LEDA  &   90  &   \\   
NGC 6172       & E      &   13.75 & LEDA  &   67  & CMZ01  \\   
NGC 6411       & E      &   12.47 & LEDA  &   53  & SMG03  \\   
NGC 6653       & E      &   13.02 & LEDA  &   66  & SMG03  \\   
NGC 6702       & E      &   12.61 & LEDA  &   66  & CMZ01  \\    
NGC 6776       & E      &   12.51 & LEDA  &   70  &   \\    
NGC 6799       & E      &   12.98 & LEDA  &   65  & RLR96, CMZ01  \\    
NGC 6849       & E/S0   &   12.93 & LEDA  &   79  & CMZ01  \\   
NGC 7330       & E      &   12.66 & NED   &   74  &   \\    
NGC 7796       & E      &   12.08 & LEDA  &   42  &   \\   
MCG-01-27-013  & E/S0   &   14.71 & LEDA  &  121  &   \\   
MCG-01-03-018  & E/S0   &   14.13 & LEDA  &   77  &   \\    
MCG-02-13-009  & E      &   13.04 & LEDA  &   73  &   \\    
MCG-03-26-030  & E/S0   &   14.26 & LEDA  &  119  &   \\   
ESO 107-G004   & E      &   12.55 & LEDA  &   39  &   \\   
ESO 153-G003   & E      &   13.66 & NED   &   84  &   \\    
ESO 194-G021   & E/S0   &   13.34 & LEDA  &   41  &   \\   
ESO 218-G002   & E      &   13.66 & LEDA  &   54  &   \\   
ESO 318-G021   & E      &   13.24 & LEDA  &   62  & RLR96  \\   
ESO 462-G015   & E      &   12.46 & LEDA  &   77  &   \\   
IC  1211       & E      &   13.48 & LEDA  &   78  &   \\   
UGC 1735       & E      &   13.42 & NED   &  109  &   \\    
UGC 2328       & E      &   13.11 & NED   &   68  &   \\    
\hline 
\end{tabular}
\end{center}
Notes: Hubble types are from LEDA. 
$B$$_T$ is the total $B$ magnitude corrected for Galactic extinction from 
Schlegel et al. (1998). 
The sources of the $B$ magnitude are listed in the next column. 
Distances obtained using the Virgo corrected 
recession velocities (from LEDA) with H$_0$ = 75 km s$^{-1}$ Mpc$^{-1}$. 
Some galaxies were listed in previous samples such as: Reduzzi, Longhetti \& 
Rampazzo (1996, RLR96), Colbert, Mulchaey \& Zabludoff (2001, CMZ01) and 
Smith, Martinez \& Graham (2003, SMG03).
 \end{minipage}
\end{table*}

\begin{table*}

\begin{center}
\begin{tabular}{llclcl}
\multicolumn{6}{l}{\bf Table 2. \small The  comparison sample.}\\
\hline
  Galaxy & Type & $B$$_T$ & Magnitude & Dist. & Environment\\ 
         &      & (mag) &  source   & (Mpc) &  \\
\hline 
IC 4320  & S0   & 13.85 & NED  & 89 & Isolated pair \\
NGC 3528 & S0   & 12.88 & LEDA & 48 & Group \\
NGC 3557 & E    & 10.79 & NED  & 38 & Group \\
NGC 4697 & E    & 10.02 & LEDA & 17 & Group \\
NGC 5266 & E/S0 & 11.17 & LEDA & 38 & Group \\

\hline 
\end{tabular}
\end{center}
Notes: Same as Table 1 for the comparison sample. The table includes the 
environment of each galaxy.
\end{table*}

\section{Sample Selection}
Our sample of candidate isolated early-type galaxies was taken from the 
LEDA. This catalogue contains information on
$\sim 100,000$ galaxies, of which $\sim 40,000$ have enough information
recorded to be of use in this work. From this sample, we selected galaxies 
which satisfied the following criteria:
\begin{itemize}
\item Morphological type $T\leq-3$, i.e. early-type.
\item Virgo corrected recession velocity $V\leq9,000$ km s$^{-1}$, i.e. 
within 120 Mpc.
\item Apparent magnitude $B$$ \leq 14.0$.
\item Galaxy not listed as a member of a Lyon Galaxy Group (Garcia 1993).
\end{itemize}
 
\hspace{5mm}
The restrictions on apparent magnitude and recession velocity were imposed
to minimize the effect of incompleteness in the catalogue. The LEDA
catalogue is known to be 90\% complete at $B=14.5$
(Amendola et al. 1997),
so our sample should be close to being 100\% statistically complete.
The selection process produced 330 galaxies which could be considered as 
potential candidates. These were compared to the rest of the catalogue and 
accepted as being isolated if they had no neighbours which were within:
 \begin{itemize}
\item 700 km s$^{-1}$ in recession velocity,
\item 0.67 Mpc in the plane of the sky, and
\item 2.0 $B$-band magnitudes of the isolated galaxy.
\end{itemize}
These criteria were imposed to ensure that the galaxies did not lie in
groups or clusters and that any neighbouring galaxies were
too small and too distant to have any significant effect on the primary galaxy.
 
To check the results of this process, all galaxies were
compared to the NED and the DSS. A NED search in the area within 0.67 Mpc
of the galaxy identified galaxies which are not listed in LEDA. We
also examined the DSS images for galaxies of similar brightness to the
target which are not listed in either catalogue. This process produced
36 candidate isolated early-type galaxies. Basic data for this sample are 
listed in Table 1. In this paper we present imaging for 10 of these galaxies. 
We have also included IC 4320 which is a part of an interacting isolated 
galaxy pair (with the spiral ESO509-G100).
They have a $B$-magnitude difference 
of 0.88 and recession velocity difference of 260 km s$^{-1}$. The projected 
distance between them is 8.2 arcmin or 212 kpc (Soares et al. 1995). 
We also include imaging of a few galaxy groups to act 
as a comparison to our isolated galaxies. These are the NGC 3557 
(LGG 229), NGC 4697 (LGG 314) and NGC 5266 (LGG 356) groups. 
In our comparison group sample, we also 
have included NGC 3528. Although it is not in the 
Garcia (1993) group catalogue, it has a luminous late-type galaxy (NGC 3529) 
at a projected distance of 5 arcmin or 70 kpc
with $B$-band magnitude difference of 1 and  
recession velocity difference of 77 km s$^{-1}$.
There is also a galaxy group (USGC S160)
located at a projected distance of 14.8 arcmin or 207 kpc from NGC 3528, 
with a recession velocity difference of 30 km/s. In addition, there are many 
intermediate luminosity galaxies in the field of NGC 3528 with no published 
magnitude and/or velocity.
The basic data for our comparison sample are summarized in Table 2.

\section{Observations and data reduction}\label{sec:obs}

For all galaxies in tables 2 and 3, $B$ and $R$-band images were obtained using the Wide Field 
Imager (WFI) on the 3.9-m Anglo-Australian Telescope (AAT) on 2002 February 
17th-19th. WFI is
an 8 CCD imaging mosaic, of $2048 \times 4096$ pixels thinned
back-illuminated CCDs with a pixel scale of $0.229^{''}$ pix$^{-1}$ giving 
a field-of-view of $30.6 \times 30.6$ sq. 
arcmins. Over the three nights there was some partial cloud and
typical seeing conditions of about $2.0^{''}$ in $B$ and $1.5^{''}$ in
$R$. The exposure times for the galaxies and the number of 
observations are listed in Table 3. In addition to the galaxies, 
several standard star fields from Landolt (1992) were obtained. These 
consisted of 10 sec exposures for both $B$ and $R$-bands.

All galaxy and standard star images were reduced in an identical manner
using IRAF tasks. Six of the WFI CCD's have one or two
bad columns which were corrected by
interpolating the pixel values on both sides of the bad
column. Data reduction included subtraction of the overscan regions.
 These regions were then trimmed before
correcting the frames by master bias subtraction, dark frame
subtraction and flat fielding using combined dome flats. The final 
images are flat to $\le 2\%$. 
We multiplicatively corrected for the different gains between CCDs using
measurements of the mean sky level in each CCD, relative to the
`best' CCD (i.e. CCD6 which is cosmetically the cleanest).

After determining an optimal aperture size of 10 pixels ($\sim 2.3 ^{"}$), 
based on a curve-of-growth type analysis, raw magnitudes of between 7 and 
25 stars were obtained for each filter using the IRAF task QPHOT.
The zero-point for each filter was determined by a
simple linear fit to the stellar raw magnitude versus their
colours published by Landolt (1992) and corrected for airmass. The
atmospheric extinction coefficients used were $k_B=0.22$ and $k_R=0.08$ 
mag/airmass. 
 The final photometric zero-points are Z$_B=25.19\pm 0.06$ and
 Z$_R=25.93\pm 0.04$.

\begin{table}
\begin{center}
\renewcommand{\arraystretch}{1.0}
\begin{tabular}{lccl}
\multicolumn{4}{c}{\bf Table 3. \small Observational data of the 
combined sample.}\\
\hline
Galaxy & A$_B$ & A$_R$  & Exp. Time  \\ 
  & (mag) & (mag) & (sec) \\
\hline 

NGC 1045 & 0.18 & 0.11 & ($B$) 1$\times$240 \\
         &      &      & ($R$) 1$\times$120  \\ 

NGC 1132 & 0.27 & 0.17 & ($B$) 1$\times$240 \\ 
         &      &      & ($R$) 1$\times$120 \\ 

NGC 2110 & 1.62 & 1.00 & ($B$) 1$\times$240 \\
         &      &      & ($R$) 1$\times$120 \\ 

NGC 2865 & 0.36 & 0.22 & ($B$) 1$\times$240 \\
         &      &      & ($R$) 1$\times$120 \\ 

NGC 4240 & 0.23 & 0.15 & ($B$) 1$\times$240 \\
         &      &      & ($R$) 1$\times$120 \\ 

NGC 6172 & 0.51 & 0.31 & ($B$) 1$\times$240  \\ 
         &      &      & ($R$) 1$\times$120 \\ 

MCG-01-27-013 & 0.19 & 0.12 & ($B$) 2$\times$240 \\
              &      &      & ($R$) 2$\times$120 \\ 

MCG-03-26-030 & 0.22 & 0.14 & ($B$) 1$\times$240  \\
              &      &      & ($R$) 1$\times$120 \\ 

ESO218-G002 & 0.75 & 0.46 & ($B$) 2$\times$120 \\
            &      &      & ($R$) 2$\times$60  \\ 

ESO318-G021 & 0.35 & 0.22 & ($B$) 1$\times$120 \\  
            &      &      & ($R$) 1$\times$60  \\ 
 \\
IC 4320     & 0.26 & 0.16 & ($B$) 1$\times$240  \\
            &      &      & ($R$) 1$\times$120  \\ 
 \\
NGC 3528    & 0.17 & 0.11 & ($B$) 1$\times$240 \\
            &      &      & ($R$) 1$\times$120 \\           

NGC 3557    & 0.43 & 0.26 & ($B$) 1$\times$60, 3$\times$600 \\
            &      &      & ($R$) 1$\times$60, 3$\times$420 \\ 

NGC 4697    & 0.13 & 0.08 & ($B$) 1$\times$60, 3$\times$600 \\  
            &      &      & ($R$) 1$\times$60, 3$\times$420 \\ 

NGC 5266    & 0.38 & 0.24 & ($B$) 2$\times$240 \\
            &      &      & ($R$) 2$\times$120 \\ 
\hline
\end{tabular}
\end{center}
Notes: The Galactic extinctions A$_B$  and A$_R$ are from Schlegel et al. 
(1998). The table has been divided into three sections based on environment.
\end{table}

\section{Measuring Photometric Parameters}\label{sec:obs}
\subsection{Magnitudes and colours}
The QPHOT task in IRAF was used to obtain total magnitudes 
for the primary galaxies in our sample. The sky was generally 
computed in an annulus of radius 1000 
pixels and width of 50 pixels. Total magnitudes were derived by fitting 
a curve-of-growth to galaxy aperture magnitudes from 
3$\times$ the seeing radius to large radii.  
Finally, the total magnitudes $B$ and $R$ were corrected for 
Galactic extinction using values from Schlegel, Finkbeiner \& Davis 
(1998) as given in Table 3. 

For ESO318-G021 and NGC 4240, the $B$ frame is dominated by the light from a 
close bright star which prevented us measuring total $B$ magnitudes. 
Considering the typical colour for galaxies in 
the present sample and the $B$ magnitude quoted by LEDA, then our measured $R$ 
magnitude for both of these two galaxies seems reasonable. A mean total 
($B$--$R$) colour of $1.46\pm0.12$ was obtained for the 13 remaining galaxies.
Our measured total $B$ and $R$ magnitudes, corrected for Galactic extinction, 
are given in Table 4. Fig. 1 shows the good agreement between our $B$ 
magnitudes compared to the published values from LEDA.

\begin{figure}
\centerline{\psfig{figure=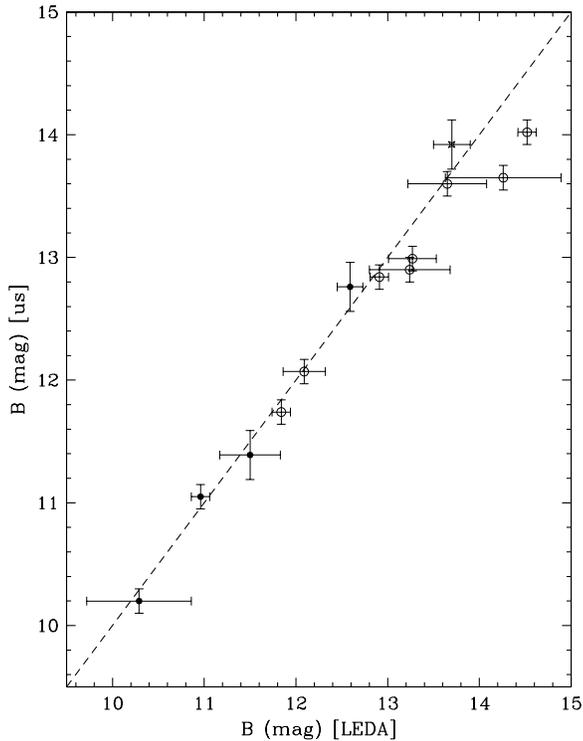,width=0.45\textwidth}}
\caption
{
Comparison between our measured $B$ magnitudes and 
the published values in LEDA. Filled and open circles represent group and 
isolated galaxies respectively, while the star is the primary galaxy
of an isolated pair (IC 4320). Our measurements show good consistency
with values from LEDA.
}
\label{plot}
\end{figure}
\begin{figure}
\centerline{\psfig{figure=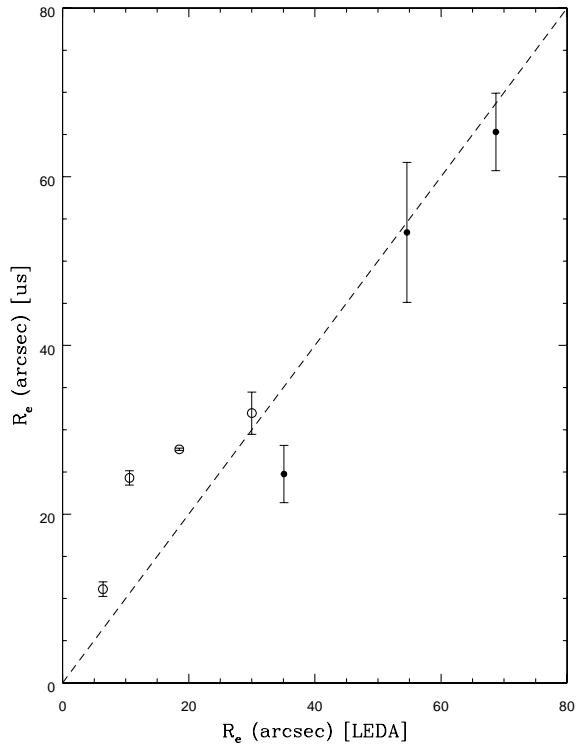,width=0.45\textwidth}}
\caption
{
Comparison between the effective radius R$_e$ measured from our $R$-band 
images and the published values in LEDA. Error bars represent 
the variation during the fitting. Generally, the agreement is good.
Symbols as in Fig. 1. 
}
\label{plot}
\end{figure}

\begin{figure}
\centerline{\psfig{figure=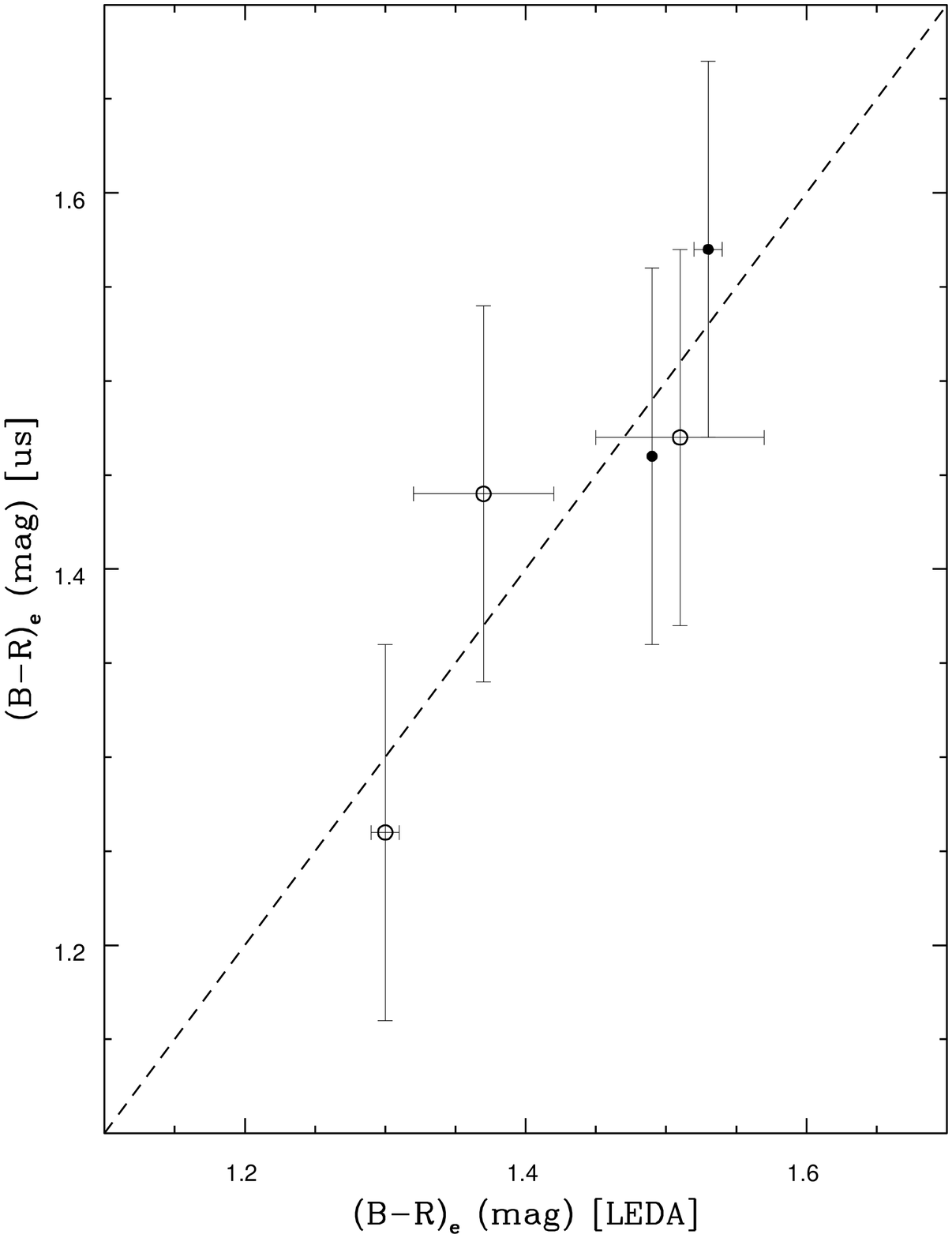,width=0.45\textwidth}}
\caption
{
Comparison between our measured colour at the effective radius ($B$--$R$)$_e$ 
and the published values in LEDA. Generally, the agreement is very good. 
Symbols as in Fig. 1.
}
\label{plot}
\end{figure}

\subsection{Modelling and surface brightness profile fitting}

Visual inspection of our CCD images of the 15 galaxies in our sample 
confirms their early-type morphology as published in LEDA.
The ISOPHOTE package in IRAF was used to fit a smooth elliptical model to the 
galaxy image (e.g. Forbes \& Thompson 1992). 
In this package the isophotes are fit using an iterative method described  
by Jedrzejewski (1987). Before 
modelling the galaxy, the sky value was determined and subtracted. The program 
starts at a small radius 
and increases to large radii in a geometrical progression. The galaxy
centre, position angle and ellipticity were allowed to vary. 

Since our sample consists of early-type galaxies with absolute magnitudes in 
the range $-19.2>$ M$_{B}> -22.1$ their surface brightness profiles can 
be well fit by a de Vaucouleurs R$^{1/4}$ law (Prugniel \& Simien 1997). By 
fitting the surface brightness profiles we derived the effective radii 
(R$_{e}$) and mean surface brightnesses ($<\mu_{e}>$) within R$_{e}$. To avoid 
resolution effects, the fitting was applied from a radius of about $3\times$ 
the seeing to large radii. In the case of NGC 5266 there is a strong dust ring 
around the galaxy (see also Goudfrooij et al. 1994a). This prevented us from
deriving a reliable surface brightness profile and parameters in the $B$-band. 
However for the $R$-band, only a small region of the profile was affected. 
This was masked and excluded from the fit. We were also unable to measure 
reliable profiles in the $B$ and $R$-bands for NGC 4240 and the $B$-band for 
ESO318-G021, due to the bright nearby stars.

Using the ellipticity ($\epsilon$) and the effective semi-major axis (a$_e$) 
from the galaxy model, the effective 
radius R$_e$ was calculated, i.e. R$_e= a_e \sqrt{1-\epsilon}$. The values of 
R$_e$, and its error estimate based on the variation during the fitting 
procedure, are given in Table 4. There is generally good agreement between the 
$B$ and $R$-band effective radii determinations. 
Fig. 2 shows that our measured values of R$_e$ from the $R$-band images and the
corresponding values from LEDA, for four isolated and three group galaxies for 
which data are available, are comparable. 

The ($B$--$R$)$_e$ colours at R$_e$ (Table 4) are obtained from the models 
using the difference between the $B$ and $R$-band surface brightness profiles. 
The exact radius used does not strongly affect the ($B$--$R$)$_e$ colour 
because of the shallow slope in the colour gradient. 
Fig. 3 shows good agreement between our derived colours and those given in 
LEDA, when available. 

The mean surface brightness within R$_e$ in each filter is 
calculated as: $<\mu_{e}>=2.5 \log[\pi$R$_e^{2}] +$ m(R$_e$), where
m(R$_e$) is the enclosed magnitude within R$_e$ estimated from the model. 
Table 5 lists the mean surface brightnesses within the effective radius in the 
$B$ and $R$-bands. 

\begin{table*}
\begin{center}

\begin{tabular}{lcccccccccc}
\multicolumn{11}{c}{\bf Table 4. \small Measured photometric and size 
parameters.}\\
\hline
Galaxy & $B$ & $\pm$ & $R$ & $\pm$ & ($B$--$R$)$_{e}$ & $\pm$ &
R$_e$($B$) & $\pm$ & R$_e$($R$) & $\pm$ \\
   & (mag) &  & (mag) &  & (mag) &  & (arcsec) &  & (arcsec) &  \\
\hline

NGC 1045 & 13.0 & 0.1 & 11.4 & 0.1 & 1.63 & 0.14 & 15.7 & 4.2  & 16.7 & 1.6 \\ 
NGC 1132 & 12.9 & 0.1 & 11.6 & 0.1 & 1.70 & 0.14 & 29.8 & 3.6  & 32.0 & 3.5 \\ 
NGC 2110 & 11.7 & 0.1 & 10.3 & 0.1 & 1.47 & 0.14 & 28.3 & 1.4  & 24.3 & 1.2 \\ 
NGC 2865 & 12.1 & 0.1 & 10.8 & 0.1 & 1.26 & 0.14 & 28.3 & 0.8  & 27.7 & 0.2 \\
NGC 4240 & - & - & 11.3 & 0.1 & - & - & - & -  & - & - \\ 
NGC 6172 & 13.6 & 0.1 & 12.3 & 0.1 & 1.44 & 0.14 & 12.4 & 1.9 & 11.1 & 1.2 \\ 
MCG-01-27-013  & 14.0 & 0.1 & 12.5 & 0.1 & 1.58 & 0.14 & 16.1 & 2.1 & 13.8 & 
0.9 \\ 
MCG-03-26-030  & 13.7 & 0.1 & 12.1 & 0.1 & 1.63 & 0.14 & 14.3 & 0.6 & 15.6 & 
2.5 \\  
ESO218-G002    & 12.8 & 0.1 & 11.4 & 0.1 & 1.58 & 0.14 & 22.0 & 6.6 & 20.1  & 
5.0 \\
ESO318-G021    & - & - & 12.1 & 0.1 & - & - & - & - & 21.7 & 3.2 \\ 
 \\
IC 4320  & 13.9 & 0.2 & 12.3 & 0.1 & 1.59 & 0.22 & 27.6 & 9.5  & 22.4 & 4.5 \\
 \\ 
NGC 3528 & 12.8 & 0.2 & 11.2 & 0.1 & 1.63 & 0.22 & 43.2 & 15. & 45.6 & 0.9 \\ 
NGC 3557 & 11.1 & 0.1 & 9.6  & 0.1 & 1.57 & 0.14 & 29.7 & 4.2 & 24.8 & 4.8 \\
NGC 4697 & 10.2 & 0.1 & 8.8  & 0.1 & 1.46 & 0.14 & 86.9 & 9.9 & 65.3 & 6.5 \\
NGC 5266 & 11.4 & 0.2 & 9.9  & 0.1 & 1.31 & 0.22 &  -   &  -  & 53.4 & 11.7 \\ 
\hline
\end{tabular}
\end{center}
Notes: The table lists total magnitudes, colour at the effective radius and the
effective radius itself. 
\end{table*}

\begin{table*}
\begin{center}
\begin{tabular}{lcccccccccccc}
\multicolumn{13}{c}{\bf Table 5. \small Derived parameters.}\\
\hline
Galaxy & M$_B$ & $\pm$ & M$_R$ & $\pm$ & $<\mu_{e}>_B$ & $\pm$ & 
$<\mu_{e}>_R$ & $\pm$ & $\log$(R$_e)_B$ & $\pm$ & 
$\log$(R$_e)_R$ & $\pm$ \\
   & (mag) &  & (mag) &   & (mag/sq.'') &  & (mag/sq.'') &  & (kpc) &  & (kpc) &  \\
\hline
NGC 1045       & -20.9 & 0.1 & -22.5 & 0.1 & 20.8 & 0.1 & 19.4 & 0.1 & 0.66 
& 0.10  & 0.69 & 0.04\\
NGC 1132       & -21.4 & 0.1 & -23.2 & 0.1 & 22.5 & 0.1 & 21.0 & 0.1 & 1.12 
& 0.05  & 1.14 & 0.06\\ 
NGC 2110       & -20.5 & 0.1 & -21.9 & 0.1 & 20.7 & 0.1 & 19.0 & 0.1 & 0.59 
& 0.02  & 0.52 & 0.02\\ 
NGC 2865       & -20.6 & 0.1 & -21.9 & 0.1 & 20.8 & 0.1 & 19.5 & 0.1 & 0.68 
& 0.01  & 0.67 & 0.01\\
NGC 4240       &   -   &  -  & -20.8 & 0.1 & -    & -   &   -  &  -  &  -   
&  -   & - & -\\ 
NGC 6172       & -20.5 & 0.1 & -21.8 & 0.1 & 20.8 & 0.1 & 19.1 & 0.1 & 0.61 
& 0.06  & 0.56 & 0.06\\ 
MCG-01-27-013  & -21.4 & 0.1 & -22.9 & 0.1 & 22.0 & 0.1 & 20.1 & 0.1 & 0.97 
& 0.05  & 0.91 & 0.03\\ 
MCG-03-26-030  & -21.7 & 0.1 & -23.3 & 0.1 & 21.4 & 0.1 & 19.9 & 0.1 & 0.92 
& 0.02  & 0.95 & 0.07\\  
ESO218-G002    & -20.9 & 0.1 & -22.3 & 0.1 & 21.5 & 0.1 & 19.8 & 0.1 & 0.76 
& 0.11  & 0.72 & 0.10\\
ESO318-G021    &   -   &  -  & -21.9 & 0.1 &  -   &  -  & 20.6 & 0.2 &   -  
&   -  &  0.81  &  0.06 \\ 
 \\					   				
IC 4320        & -20.9 & 0.2 & -22.5 & 0.1 & 22.7 & 0.2 & 20.8 & 0.1 & 1.08 
& 0.13  & 0.99 & 0.08\\			   				
 \\ 
NGC 3528       & -20.6 & 0.2 & -22.2 & 0.1 & 22.4 & 0.2 & 20.9 & 0.1 & 1.00 
& 0.13  & 1.03 & 0.01\\ 
NGC 3557       & -21.8 & 0.1 & -23.3 & 0.1 & 20.5 & 0.1 & 18.6 & 0.1 & 0.74 
& 0.06  & 0.66 & 0.08\\
NGC 4697       & -21.0 & 0.1 & -22.4 & 0.1 & 21.4 & 0.1 & 19.5 & 0.1 & 0.86 
& 0.05  & 0.73 & 0.04\\
NGC 5266       & -21.5 & 0.2 & -23.0 & 0.1 &   -  &  -  & 20.1 & 0.1 &  -   
&   -   & 0.99 & 0.09\\ 
\hline
\end{tabular}
\end{center}
Notes: The table lists absolute magnitudes, mean surface brightness within the 
effective radius and the effective radius in kpc.
\end{table*}

\subsection{Isophotal shape parameters}
The modelling process fits elliptical isophotes to the galaxy images and 
measures their isophotal shape parameters. Figures 4-7 show 
the radial profile of these parameters in the $R$-band for the fourteen 
galaxies which could be successfully modelled (i.e. excluding NGC 4240).
These isophotal shape parameters help  define the morphology of the galaxy.
For example, the 4$^{th}$-order cosine term of the Fourier series is a useful 
parameter to 
express the deviation from a perfect ellipse due to the presence of additional 
light. An excess of light along the major and/or minor axes (called disky) is 
indicated by positive values, while negative values indicate excess light at 
$45^\circ$ with respect to these axes (called boxy). 

To quantify the $4^{th}$-order cosine parameter we applied a 
technique similar to that of Bender et 
al. (1989). When the fourth-order cosine profile shows a peak or a 
minimum, then the maximum amplitude is taken, while in case of a monotonically 
changing profile we use the value at R$_e$. In the case of more complicated 
profiles indicated by a radial change from disky to boxy (or vice 
versa), such galaxies are classified as irregular (irr).
Table 6 lists the $4^{th}$-order cosine values normalised to the
semi-major axis. It also lists the difference between the maximum and
minimum values of the 
position angle for each galaxy after avoiding the inner, seeing-affected 
regions.

\begin{figure}
\caption
{
The $R$-band (open circles) surface brightness, ellipticity, position angle 
and 4$^{th}$-order cosine profiles. The upper panel for each galaxy shows
also the $B$-band surface brightness (solid circles). The name of each
galaxy is written above each set of four panels. [See fig4to7.tar]
}
\end{figure}
\begin{figure}
\caption
{
Same as Fig. 4. [See fig4to7.tar]
}
\end{figure}

\begin{figure}
\caption
{
Same as Fig. 4. [See fig4to7.tar]
}
\label{plot}
\end{figure}
\begin{figure}
\caption
{
Same as Fig. 4. [See fig4to7.tar]
}
\label{plot}
\end{figure}

\begin{figure}

\caption
{
Residual images of galaxies (with filters given in brackets): NGC 1045($R$), 
NGC 1132($R$), NGC 2110($R$), NGC 2865($R$), 
NGC 6172($R$), MCG-01-27-013($R$), MCG-03-26-030($R$), ESO218-G002($R$) and 
ESO318-G021($R$) (left to right, top to bottom).
Dust regions can be seen as bright features and extra light as 
dark features. All images have the same size of 188 $\times$ 155 sq. 
arcsec and oriented as North up and East to the left. [See fig8and9.tar]
}
\label{plot}
\end{figure}

\begin{figure}

\caption
{
 Residual images of IC 4320($B$), NGC 3528($B$), NGC 3557($R$), NGC 4697($R$) 
and NGC 5266($B$) (left to right, top to bottom). 
Orientations and sizes as in Fig. 8. [See fig8and9.tar]
}
\label{plot}
\end{figure}
\begin{figure}
\caption
{
The size-magnitude relation for detected galaxies in the field of (a) the 
isolated galaxy ESO218-G002 and (b) the NGC 3528 group. 
The SExtractor parameters A\_IMAG and MAG\_BEST were used as a measure 
of size and magnitude respectively. The vertical dotted line represents our 
100\% completeness limit of R $=18.75$. For ESO218-G002 the brightest 
galaxy in the field is $\sim 4$ mags fainter and $\sim 1/3$ the size of 
ESO218-G002 itself. [See fig10.gif]
}
\label{plot}
\end{figure}

\begin{table}
\begin{center}
\renewcommand{\arraystretch}{1.0}
\begin{tabular}{lcc}
\multicolumn{3}{c}{\bf Table 6. \small Isophotal shape parameters. }\\
\hline
Galaxy & $4^{th}$ cosine \% & $\Delta$PA ($^{\circ}$) \\
\hline

NGC 1045  & -3.5 & 0  \\

NGC 1132  & -1.5 & 10 \\

NGC 2110  & irr & 12  \\

NGC 2865  & +1.5 & 14 \\

NGC 6172  & irr & 25  \\ 

MCG-01-27-013 & +1.0 & 14 \\

MCG-03-26-030& +2.0 & 2 \\

ESO218-G002 & 0  & 70 \\

ESO318-G021 & 0 & 8 \\
 \\
IC 4320 & irr  & 30  \\
 \\ 
NGC 3528 & +3.2 & 10 \\

NGC 3557 & 0 & 0 \\

NGC 4697 & +2.0 & 2 \\

NGC 5266 & irr & 8 \\

\hline  
\end{tabular} 
\end{center}
Notes: The percentage $4^{th}$ cosine term of the Fourier series is normalised 
to the semi-major axis. The difference between the maximum and minimum 
position angles are quoted for radii greater than 10 arcsec.
\end{table}

\subsection{Residual images}
Another way to study the morphology or fine structure of an elliptical galaxy 
is by examining 
the residual image, i.e. the original image minus the galaxy model. Using the 
BMODEL task in IRAF we created a 2D model of the galaxy. This model represents 
the smooth elliptical structure of each galaxy. Subtracting 
this model from the original image gives the residual image which better 
reveals any fine 
structure (e.g. dust, shells, tidal tails, boxy or disky 
structure, etc.) hidden underneath the dominant elliptical light of the galaxy.
Figures 8 and 9 display the residual images of the sample galaxies (excluding 
NGC 4240).

\section{Faint Galaxy Detection}
As our images cover a field-of-view of hundreds of square kiloparsecs 
surrounding each galaxy, they allow us to probe the distribution of 
galaxies in their fields.
With the strict isolation criteria used to select galaxies in this study, the 
area surrounding the primary galaxy will contain no bright galaxies (within 
$\sim2$ mags) but may still contain many faint ones. The detection of these 
faint galaxies was performed using SExtractor version 2.3 (Bertin \& Arnouts
 1996). From the deeper $R$-band images,  
all objects with more than 10 connected pixels that are 3$\sigma$ above the 
sky background were identified. To estimate the 
background level of the images, as well as the RMS noise in the background, we 
set the mesh size to 200 pixels. This value of the mesh size 
was found to be suitable for extended object detection. The image was then 
smoothed with a median filter of $10\times10$ pixels to remove any 
fluctuations resulting from bright or extended objects. 
The photometric parameter MAG\_BEST was used as a measure of the total 
magnitude. A comparison between MAG\_BEST magnitudes and the corresponding 
total magnitudes measured with the IRAF task QPHOT showed good agreement.

For star-galaxy separation, we used the CLASS\_STAR parameter (ICLASS) defined 
by SExtractor. Running SExtractor on the 60 sec and the combined (1320 sec) 
$R$ band mosaic frames of the NGC 3557 group, we obtained two sets of 
detected bright objects, i.e. stars and 
galaxies with ICLASS parameters of 1 and 0 respectively. For faint objects 
(R $\geqslant 18$), ICLASS values ranged from 0 to 1. 
Visual inspectation of the 60 sec and combined images revealed that an ICLASS 
value of 0.09 provided a good separation between resolved galaxies 
and stellar-like objects. Thus objects with ICLASS $> 0.09$ were removed from 
all of our object lists. After displaying all the remaining detections, we 
removed the small number of obvious mis-identifications (e.g. bad columns, 
diffraction spikes, halos of bright stars, etc.). Finally, for the isolated 
galaxies, we searched the NED database and excluded any of the detected faint 
galaxies that have a published velocity difference of greater than 
700 km s$^{-1}$
from the parent galaxy. This resulted in only a handful of galaxies being 
removed for the isolated galaxy sample.

In order to estimate our magnitude completeness limits, we compared the 60 and
420 sec to the combined (1320 sec) frame of the NGC 3557 group. 
This allows us to 
estimate our completeness as a function of exposure time.  For example, a 
typical 120 sec exposure has an estimated 100\% completeness limit of 
$R$ $=18.75$. 
Except for ESO318-G021, all other galaxies have a total exposure time of at 
least 120 sec (Table 3).

Using the SExtractor parameter A\_IMAG as a measure of galaxy size, Fig. 10
 shows the size-magnitude distribution after our selection criteria and 
visual inspection are applied. The figure shows the isolated galaxy 
ESO218-G002 and the NGC 3528 group which are at almost same distance and have 
the same exposure time of 120 sec. The vertical line indicates our 
100\% completeness limit of $R$ $=18.75$. For ESO218-G002 the brightest galaxy 
in the field is $\sim4$ mag fainter and $\sim 1/3$ the size of ESO218-G002 
itself. As a result of the 
SExtractor process, we have a list of apparent magnitudes, projected distances 
in kpc from the primary galaxy
and angular sizes of galaxies in the field of the primary galaxy. 

\section{Results and discussion}
In this section we first discuss our results about the primary galaxy sample 
and then the results related to the detected galaxies in their fields.

\subsection{Colours of isolated galaxies}
Prugniel \& Heraudeau (1998), using a sample of 5,169 galaxies, measured the 
colour for S0 galaxies to be ($B$--$R$)$_e= 1.52 \pm 0.2$ and for ellipticals 
($B$--$R$)$_e= 1.57 \pm 0.2$. 
These values are consistent with the measured effective colours of our sample 
(see Table 4) which have a mean of ($B$--$R$)$_e= 1.54 \pm 0.14$. 
We note that NGC 2865 is quite blue with ($B$--$R$)$_e= 1.26\pm 0.14$.

The CMR of ellipticals was discovered by Baum 
(1959) when he studied the stellar content for a sample of elliptical 
galaxies and globular clusters. He found more luminous ellipticals to be 
redder. Further studies of cluster ellipticals have shown 
that the CMR is a linear relation with a small intrinsic scatter (e.g. 
Bower, Lucey \& Ellis 1992). 
A comparison of the observed slope and scatter of the CMR with that 
expected from theoretical models, has been used by many 
authors to constrain the formation epoch of the 
 bulk of stars in early-type cluster galaxies to be at $z>2$ (e.g. 
Bower et al. 1992; Stanford,  Eisenhardt \& Dickinson 1995; Ellis et al. 
1997; Stanford et al. 1998; Bower, Kodama \& Terlevich 1998; 
Bernardi et al. 2003).  
The constant slope out to high redshifts means that the CMR  
is generally interpreted as a relation of mass with metallicity 
(Kodama 2001); while the scatter, although small, is due to ongoing star 
formation.

In Fig. 11, we show that our isolated galaxies are reasonably well matched by 
the $B$--$R$ CMR for the Coma cluster derived by Gladders et al. (1998). 
We used total $B$--$R$ colours as they are the closest match to the aperture 
magnitudes of Gladders et al. Fixing the slope to the Gladders et al. value, 
we measure a scatter of $\sim 0.12$ mag for the isolated galaxies, which is 
similar to our photometric errors. Thus the scatter is largely intrinsic and 
consistent with the estimated low scatter found
for early-type galaxies in clusters (Bower et al. 1992; Stanford et al. 1998).

\begin{figure}
\centerline{\psfig{figure=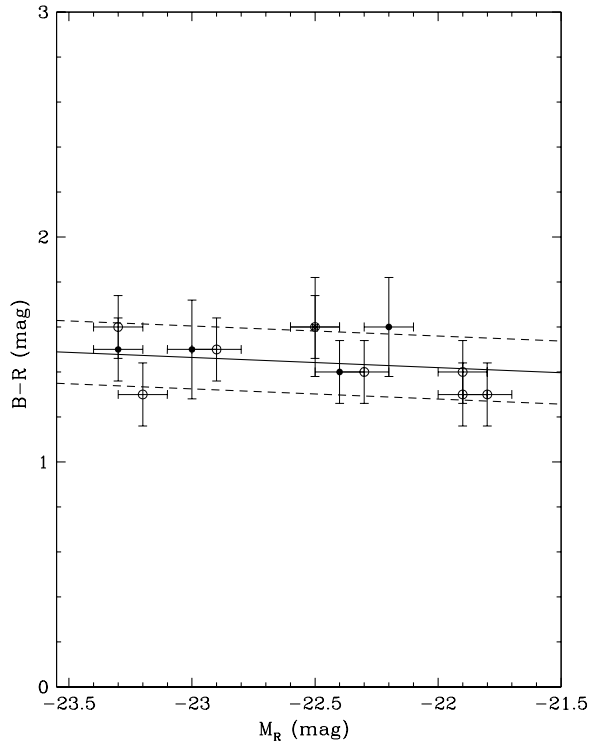,width=0.45\textwidth}}
\caption
{
Colour-magnitude relation for the sample galaxies. The solid line (Gladders et 
al. 1998) represents the CMR of elliptical galaxies in the Coma cluster, 
while the dashed line indicates the mean error in the colour measurements for 
our sample galaxies. The isolated galaxies are consistent with the Coma 
cluster CMR. Symbols as in Fig.1.
}
\label{plot}
\end{figure}
    
\subsection{Fine structure}
Many studies have found that elliptical galaxies have light 
distributions which deviate from perfect ellipticity (Lauer 1985; Bender, 
D\"oebereiner \& M\"{o}ellenhoff 1988; Franx, Illingworth \& Heckman 1989; 
Peletier et al. 1990; Bender \& M\"oellenhoff 1987; Forbes \& Thomson 1992;
Goudfrooij et al.\ 1994b). These deviations   
can be due to dust lanes or extra-light structures. Such 
morphological features could be a result of the tidal disruption of a small 
companion (Binney \& Petrou 1985; Nieto \& Bender 1989) or a major merger 
(Schweizer et al. 1990).
In a detailed study of boxyness in massive ellipticals by Nieto \& Bender 
(1989) and Bender et al. (1989), the isophotal shape showed little or
no correlation with the classic photometric parameters such as luminosity, 
effective radius or ellipticity. However, galaxies which were
radio-loud and$/$or surrounded by gaseous X-ray halos generally had boxy or 
irregular isophotes. Bender et al. also found that galaxies with 
boxy or irregular isophotes seem to have higher mass-to-light ratios 
($\sim11.5\pm0.9~ M_\odot/L_\odot$) than disky galaxies 
($\sim6.4\pm0.6~ M_\odot/L_\odot$). 

Our isolated galaxy sample reveals a range of fine structure from plumes and 
shells to dust lanes, and some galaxies with no detectable fine structure. 
Good examples of boxy and disky isophotes can be seen in NGC 1045 and NGC 4697 
respectively. Shells (in the case of NGC 2865) and dust lanes appear to affect 
the $4^{th}$-order cosine profile of many galaxies.
The presence of internal structures such as 
dust, shells, disky or boxy structure could be the result of a past 
merger (e.g. Forbes 1991; Barnes \& Hernquist 1992; Balcells 1997; 
Merluzzi 1998; Khochfar \& Burkert 2003). 
Including plumes, shells and dust, 4 out of 10 isolated 
galaxies reveal signs of a past merger or accretion event. The isolated 
galaxies ESO218-G002 and ESO318-G021 are good examples of no obvious fine 
structure. From the radial profiles (Table 6) and the residual images of each 
galaxy (Figures 8 and 9), the following features were revealed:

\noindent
\bf NGC 1045: \normalsize The isophotes are strongly boxy out to a semi-major 
axis radius of $35^{''}$. Probable dust along the semi-major and semi-minor 
axes. There is extensive extra tidal light surrounding the galaxy. This galaxy 
is a probable merger remnant.\\
\bf NGC 1132:  \normalsize No obvious features. Classified as a fossil galaxy 
by Mulchaey \& Zabludoff (1999). \\
\bf NGC 2110: \normalsize  Probable dust traces in the central region. 
Contains a seyfert nucleus (Pfefferkorn, Boller \& Rafanelli 2001).\\
\bf NGC 2865: \normalsize Extra tidal light (shells) to the east and 
north east. Probable dust in the inner region. This galaxy is a probable 
merger remnant (see also Hau, Carter \& Balcells 1999).\\
\bf NGC 4240: \normalsize Nearby bright star. No obvious features.\\
\bf NGC 6172: \normalsize No obvious features. \\
\bf MCG-01-27-013: \normalsize Weak disky structure at radius $\sim15^{''}$.\\
\bf MCG-03-26-030: \normalsize Disky beyond 10${''}$. Probable dust.\\
\bf ESO218-G002: \normalsize No obvious features. Large position angle twist 
seen. \\
\bf ESO318-G021: \normalsize Nearby bright star. No obvious features.\\
\bf IC 4320: \normalsize Strong dust lane within central $20^{''}$. Probably a 
merger remnant.\\  
\bf NGC 3528: \normalsize Strong dust lane extending from the north east to 
south west. Probably a merger remnant.\\
\bf NGC 3557: \normalsize No obvious features.\\ 
\bf NGC 4697: \normalsize Strong disky structure extending to $\sim40^{''}$.\\ 
\bf NGC 5266: \normalsize Dust ring around galaxy. Probably a merger remnant.\\
    
Thus, some isolated galaxies reveal signatures of a past interaction/merger 
while others look featureless and undisturbed. 
Zepf \& Whitmore (1993) found 50\% of their Hickson compact group elliptical 
sample to have irregular isophotes compared to $\lesssim 21\%$ in 
loose groups and clusters.
Our sample, although small, does not show a predominance of irregular 
isophotes, and appears similar to the distribution of isophotal shapes 
for loose groups and cluster ellipticals suggesting a similar frequency of 
mergers and interactions.
Our isolated galaxy sample includes 44\% (4/9) with dust and 11\% (one galaxy; 
NGC 1045) with 
tidal tails which is a higher percentage than that found by Reduzzi, 
Longhetti \& Rampazzo (1996) who quote a percentages of 24.6\% and 3.3\% 
respectively. 
Colbert et al. (2001) detected 75\% of early-type galaxies in low-density 
environments to have dust. 
The galaxies with shells in our sample show a similar percentage of 11\% (one 
galaxy; NGC 2865)
compared to 16.4\% in the sample of Reduzzi et al. (1996).
Overall, our isolated galaxies show evidence of past mergers (shells,
dust, tidal tails, etc) in 44\% of the total sample which is similar to the 
corresponding percentage in Reduzzi et al. (1996).

\subsection{Kormendy relation}
\begin{figure*}
\centerline{\psfig{figure=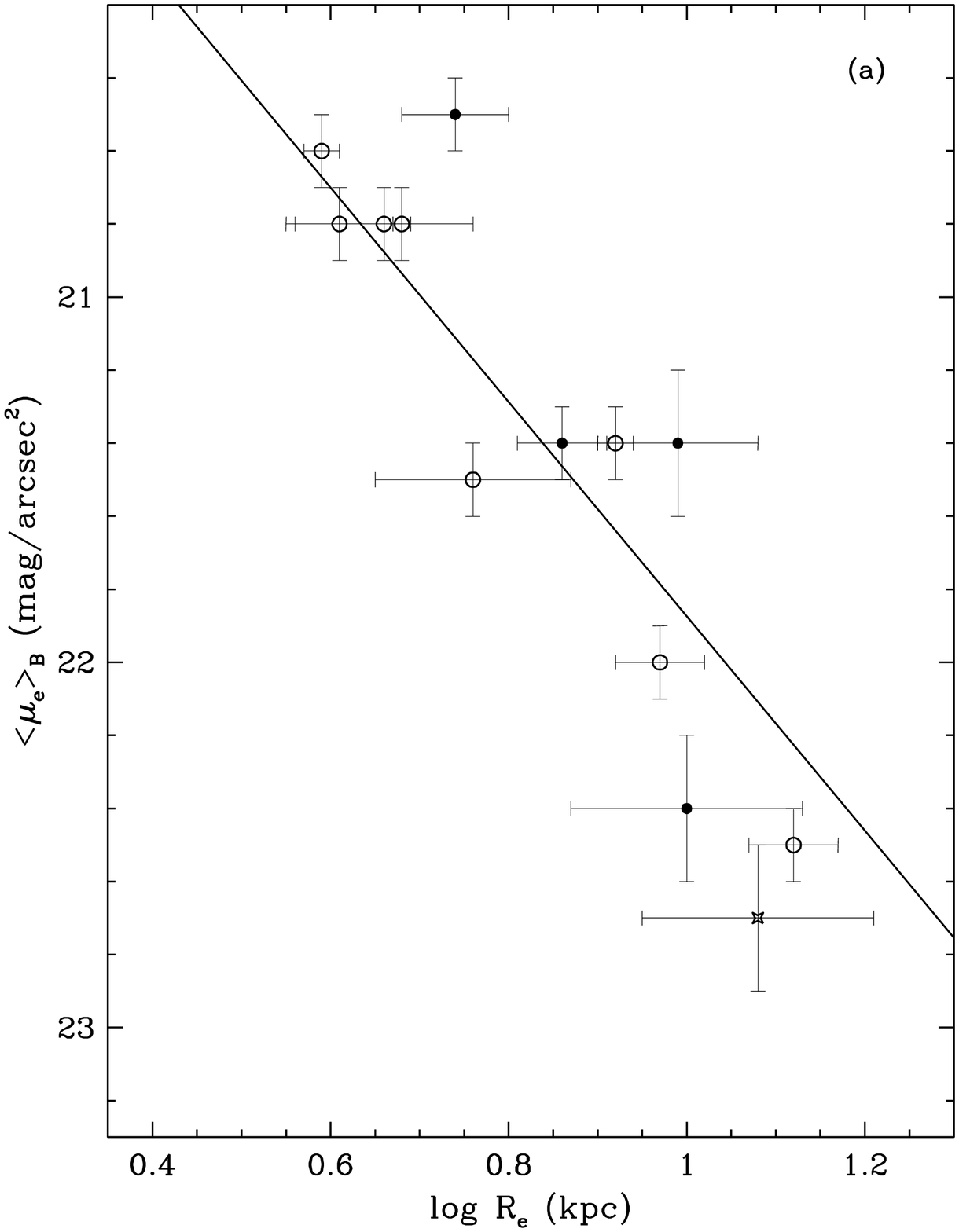,width=0.45\textwidth}
\psfig{figure=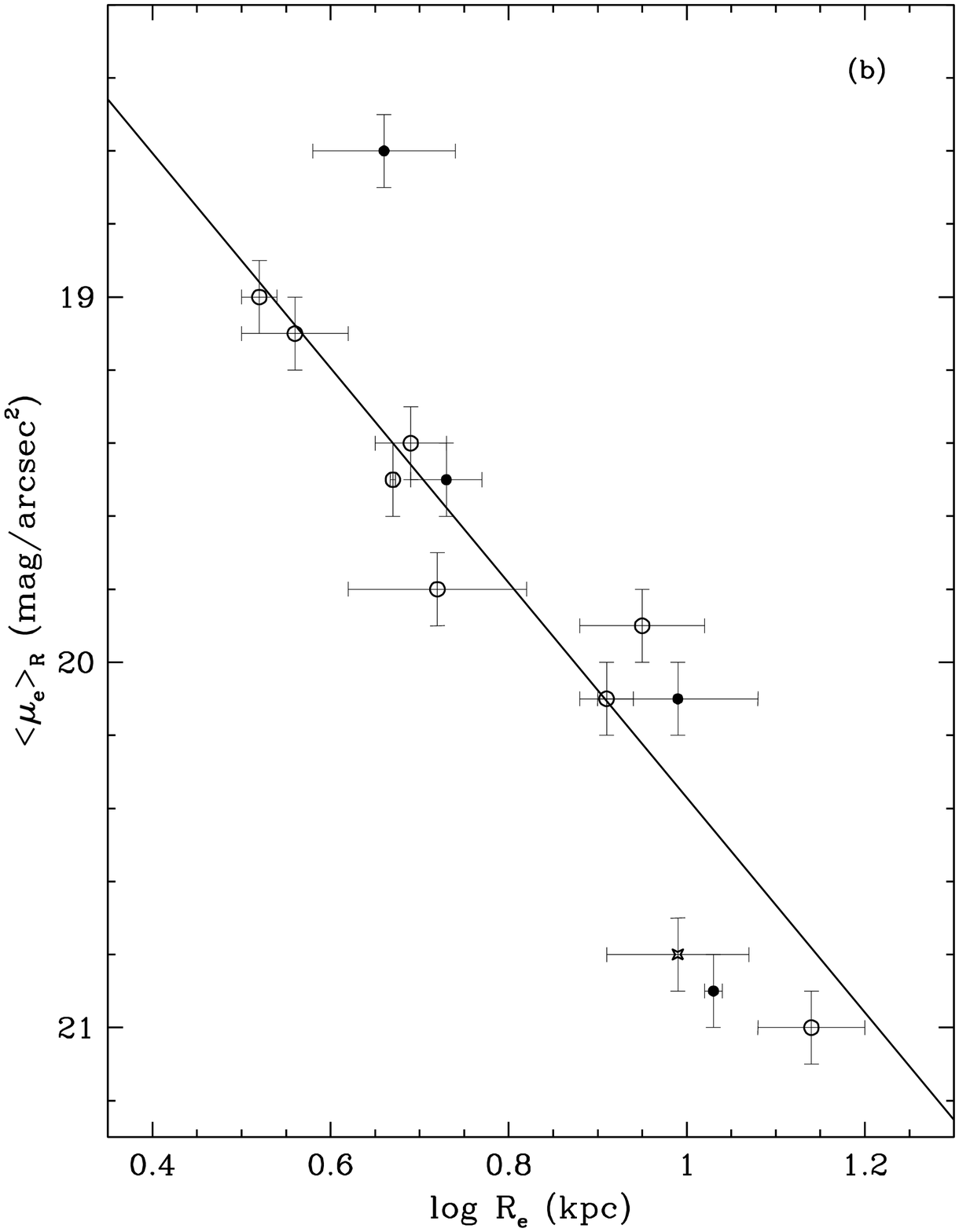,width=0.45\textwidth}}
\caption
{
The Kormendy relation between the mean surface brightness within the 
effective radius and effective radius in the $B$ and $R$-bands. The solid line
represents the original Hamabe \& Kormendy (1987) relation. The zero-point 
of the original relation has been shifted from $V$ to the $B$ and $R$-bands 
assuming typical colours of $B$--$V$ = 0.8 and $B$--$R$ = 1.5. Our isolated 
galaxies are generally consistent with the Hamabe \& Kormendy 
relation for luminous galaxies.
Symbols as in Fig. 1.
}
\label{plot}
\end{figure*}

Kormendy (1977) found that the effective radius (R$_e$) of elliptical galaxies 
is correlated with the surface brightness at R$_e$ ($\mu_e$).
Based on a larger sample, Hamabe \& Kormendy (1987) found this correlation in 
the $V$-band to be $\mu_{e} = 2.94 \log($R$_e) + 19.48$.

 Many studies have used this relation to investigate galaxy evolution with 
environment and look-back time. For example, 
Hoessel, Oegerle \& Schneider (1987) studied 372 elliptical galaxies in 97 
nearby ($z < 0.1$) Abell clusters and found that the brightest cluster 
galaxies to have a R$_e$-$\mu_{e}$ relation very similar to that in Hamabe 
\& Kormendy (1987). However they also found that the less luminous 
ellipticals in the cluster cores had a steeper relation with a slope of 4.56, 
in the sense that at a given surface brightness these   
 galaxies have smaller effective sizes. The distinction of the two 
different luminosity groups became more obvious in a later study by 
Capaccioli, Caon \& D'Onofrio (1992) with a larger sample of Virgo cluster 
galaxies and other data collected from the literature. They confirmed that, 
unlike the low luminosity galaxies, the brightest cluster galaxies fit 
perfectly on the Hamabe \& Kormendy relation.
In a recent study by Khosroshahi et al. (2004), the Kormendy relation was 
explored for early-type galaxies in groups that have a range of X-ray 
luminosities. They found that group early-type galaxies show a similar 
relation to those in clusters.
Up to redshifts of $z \sim 1.5$, elliptical galaxies in clusters show 
a similar relation to cluster galaxies in the local 
universe, with no significant change in the slope or scatter
(Ziegler et al. 1999; Waddington et al. 2002; La Barbera et al. 2003).

The distribution of our isolated galaxies in the 
$\log($R$_e)$-$<\mu_e>$ plane, 
where $<\mu_e>$ is the mean surface brightness within R$_e$, is shown in 
Fig. 12.
Using the relation between $\mu_{e}$ and $<\mu_e>$ (Graham \& Colless 
1997) and assuming typical colours of $B$--$V$ = 0.8 and $B$--$R$ = 1.5, we
reproduce the Hamabe \& Kormendy  relation in the $B$ and $R$-bands in Fig. 12.
Our isolated galaxies are generally consistent with the Hamabe \& Kormendy 
relation for luminous galaxies.

\subsection{The gravitational effect of faint companions}

Here, we have assumed that all detected faint galaxies in the field 
surrounding a galaxy lie at the same redshift as the primary galaxy 
(in reality many will be background objects) and 
have calculated their luminosity $L_R$ in the $R$-band.
For an assumed mass-to-light ratio of $M/L_R =10~ M_\odot/L_\odot$, we 
estimated the mass $M$ of each faint galaxy.
For all faint galaxies in the field of a primary galaxy, we have calculated 
the dynamical friction timescale 
using equation 7-27 in Binney \& Tremaine (1987). We assumed a velocity 
dispersion of $\sigma = 250$ km s$^{-1}$ for the galaxies in the field of the 
isolated galaxy. As galaxies with 
dynamical friction times that are significantly less than the Hubble 
time should have merged with the primary 
galaxy, and hence not be visible, their existence suggests that they may 
actually lie in the foreground or background. 
In the case of the four groups, we used the published velocities in NED to 
remove the non-group members.

The gravitational effect applied on the primary galaxy by each of the faint 
galaxies in its field is proportional to $M/R^2$, where $R$ is the 
projected distance of the faint galaxy from the primary galaxy.
Except for the faint dwarfs with short dynamical friction times and 
non-group members, we calculate the parameter $M/R^2$ for all detected 
galaxies in the field of the 10 isolated and 4 group galaxies. 
Fig. 13 shows the number distribution of the $M/R^2$ values for the 
isolated  and the group galaxies. In the case of groups, the distribution 
has been scaled up by a factor of 2.5$\times$.
Comparing the fields of the isolated and group galaxies we 
notice the absence of galaxies with $M/R^2$  greater than $\sim 4 \times 
10^6~ M_\odot/$kpc$^2$ for isolated galaxies. 
This is a further indication that our isolated galaxies are indeed 
gravitationally isolated.

\begin{figure}
\centerline{\psfig{figure=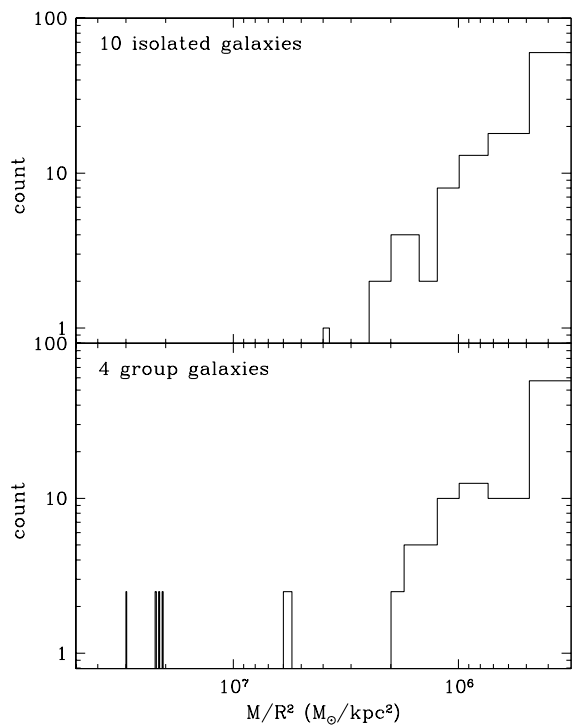,width=0.45\textwidth}}
\caption
{
Number distribution of $M/R^2$ values for the 10 isolated and 4 group 
galaxies. The group  galaxies have been scaled up in number by a factor of 
2.5$\times$ to match the isolated galaxy sample.
$M$ is the mass of the detected galaxy in solar masses and $R$ is 
the projected distance of the galaxy from the primary galaxy in kpc. 
The value $M/R^2$ represents the gravitational force applied on 
the primary galaxy by the faint galaxies in its local environment (assuming 
they lie at the same redshift). Unlike the group galaxies, the isolated 
galaxies are not influenced by the gravitational effects of faint neighbours 
with $M/R^2\gtrsim 4 \times 10^6~ M_\odot/$kpc$^2$.
}
\label{plot}
\end{figure}

\subsection{Luminosity function} 
The Luminosity Function (LF) represents the number distribution of galaxies in 
a given magnitude interval. Traditionally, this distribution is fit
with a function of the form \\
$\phi(L)=(\phi^*/L^*)(L/L^*)^\alpha$ e$^{(L/L^*)}$ \\
with $L^*$ and $\alpha$ representing a characteristic bright magnitude 
and the slope of the faint end respectively (Schechter 1976). The importance 
of the LF lies in its strong correlation with the mass function 
of the galaxy distribution, which is one of the observable tests of any 
theoretical cosmological model. It may also provide clues to 
the evolutionary history of galaxies. There have been several studies of the 
LF for high-density cluster environments 
(Trentham \& Hodgkin 2002; Mobasher et al. 2003; Christlein 
\& Zabludoff 2003) and lower density regions such as groups (Zabludoff \& 
Mulchaey 2000; Flint, Bolte \& Oliveira 2003) and the field (Blanton et al. 
2001; Blanton et al. 2003; Drory et al. 2003).

Due to the different distances to the isolated galaxies, our data 
cover different physical areas with different absolute magnitude limits (under 
the assumption that all faint galaxies lie at the same redshift as the primary 
galaxy).
To deal with this situation we divided the isolated galaxy sample into three 
distance intervals, and derived the LF separately for each (see Fig. 14).
Excluding the nearby group NGC 4697, our imaging of the other three 
groups covers a comparable area to that of the isolated galaxies. 
We note that in driving the LFs, we assume that the faint galaxies are at the 
same redshift as the primary galaxy. Spectra are required to determine the 
true redshift distribution of these galaxies.
We find that the isolated galaxies and groups have similar luminosity 
functions. They all show a faint end slope of $\sim -1.2$, and a change in the 
slope around M$_R\approx -18$. Such features are fairly common across 
a range of environments (Flint 2001; Trentham \& Tully 2002; Miles et al. 
2004). We note that 
fossil groups (Jones et al. 2000) also reveal a similar luminosity function. 
This universality of isolated galaxy luminosity functions
indicates that they are not a useful tool 
in discriminating between competing ideas for the origin
of isolated galaxies.

\begin{figure}
\centerline{\psfig{figure=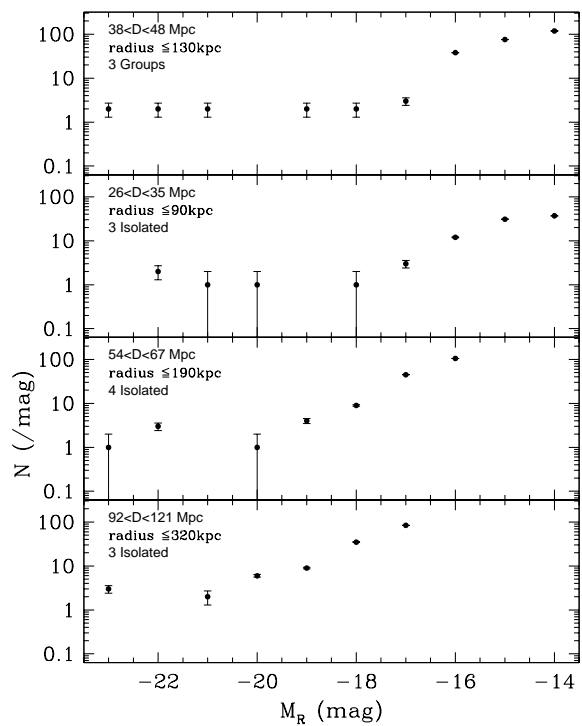,width=0.45\textwidth}}
\caption
{
Differential luminosity function in the $R$-band for faint galaxies in the 
field 
of the ten isolated galaxies divided into three categories according to their 
distance. The top panel is the LF for three groups with distances 
comparable with the nearest isolated galaxies. 
Luminosity functions show a common faint end slope of about -1.2. 
}
\label{plot}
\end{figure}

\subsection{Space number density}

To calculate the space number density of the detected faint galaxies 
surrounding the primary galaxy we divided the field into concentric annuli 
centred on each primary galaxy. The density measures in the outer annulii were 
corrected for the missing area with the largest correction being a factor of 2,
 i.e. 50\% of the annulus was covered by the WFI field-of-view. The different 
magnitude limits in each field, determined by the exposure time, give rise to 
different background density levels.

In Figures 15 and 16, we display the space number density distribution of 
faint galaxies in the field of each primary galaxy. Seven out of ten isolated 
galaxies show 
almost constant density levels at all radii, indicating that the faint galaxies
 are mostly background objects and not associated with the isolated galaxies. 
The remaining three galaxies are NGC 2110, 
NGC 2865 and NGC 4240. Both NGC 2110 
and NGC 4240 show a fairly continuous density decrease from 
the central region outwards. The outer most radial bin has a relatively 
high density suggesting that the background level has not been reached even at 
$\sim150$ kpc. NGC 2865 shows a sharp decline from a high central 
density to a background level at a radius of $\sim 60$ kpc. 

We note that these three galaxies are the closest in our sample,
and for which we could detect lower luminosity dwarfs with magnitudes down to 
M$_R \sim -13.5$. The mean faint magnitude limit that could be detected for 
the other 7 isolated galaxies is M$_R \sim -15.5$.
If we include only luminous dwarfs with M$_R \lesssim -15.5$, then 
these three galaxies show no central excess, giving similar space 
densities to those of other galaxies (Fig. 17). This indicates that 
the majority of dwarf galaxies associated with the isolated galaxies 
are less luminous than M$_R \sim -15.5$, while more luminous dwarfs are likely 
to be background objects.

 In their study of the NGC 1132 environment, Mulchaey \& Zabludoff (1999) 
measured a higher central density of dwarf galaxies, with magnitudes of  
$17.22 < $R$ < 19.22$, than we have. From their density distribution we 
calculate
the number of dwarfs within the central 80 kpc to be $\sim 30$ galaxies. 
We identify only 8 galaxies within the same area. Our magnitude limit for 
the NGC 1132 field is $R$ $=18.75$. 
This magnitude difference is not enough to explain the density distribution 
difference between the two studies. Considering all detected objects down to 
their magnitude limit of $R$ $=19.22$ and with ICLASS parameters up to 0.95, 
we find 19 objects. 
This number of detected dwarf galaxies is still less than that 
found by Mulchaey \& Zabludoff. 
This smaller number could be due to the incompleteness of our detections for 
objects fainter than $R$ $=18.75$.
 The background level in the outer region (i.e. distances $>100$ kpc) is
consistent between the two studies.

As groups are distributed over a few megaparsecs, which is beyond our 
observed field-of-view, it is not surprising that the density levels remain 
high at all radii for our four group galaxies, 
i.e. none reach the background level. For IC 4320, after excluding its pair 
galaxy ESO509-G100, shows a constant density indicating no associated galaxies.

\begin{figure}
\caption
{
Space number density of detected faint galaxies in the field of each primary 
galaxy. We show Poisson error bars. The horizontal dashed lines represent 
the estimated background density level. The name of each primary galaxy is 
indicated (N=NGC, M=MCG, and E=ESO). [See fig15to17.gif]
}
\label{plot}
\end{figure}

\begin{figure}
\caption
{
Same as Fig. 15. Note the different vertical scale for N4697. [See fig15to17.gif]
}
\label{plot}
\end{figure}
\begin{figure} 
\caption
{
Space number density of the detected dwarf galaxies in the fields of the three 
galaxies NGC 2110, NGC 2865 and NGC 4240. In this figure only luminous dwarfs 
of M$_R \lesssim -15.5$ are included. The horizontal dashed lines represent 
the estimated background density level after this magnitude selection is 
applied. The luminous dwarfs are consistent with being background objects for 
all three galaxies. [See fig15to17.gif]
}
\label{plot}
\end{figure}

\subsection{Comparison with fossil groups}

Jones et al. (2003) defined a `fossil' to be a giant isolated elliptical 
galaxy with group-like X-ray emission and luminosities of 
$L_{X,bol}\geq 10^{42} h_{50}^{-2}$ erg s$^{-1}$. Fossils are required to be 
$\geq 2.0$ $R$ magnitudes brighter than the second brightest galaxy in 
the system  within half the projected virial radius.  
According to the hierarchical model of galaxy formation these elliptical 
galaxies are believed to be the end result of merging many small galaxies 
of a X-ray luminous group (Ponman et al. 1994; Jones, Ponman \& Forbes 2000; 
Jones et al. 2003). 

We have searched the literature for X-ray observations of our sample 
galaxies, finding  
only NGC 2865, NGC 3557 and NGC 4697 (O'Sullivan, Forbes \& Ponman 2001), 
NGC 2110 (Bradt et al. 1978; Pfefferkorn, Boller \&  Rafanelli 2001),
NGC 1132 (Mulchaey \& Zabludoff 1999) and NGC 3528 and NGC 5266 (Horner \& 
Goudfrooij 2004)  to have published X-ray luminosities. 
Only NGC 1132 has an X-ray luminosity 
($L_X=2.5\times10^{42} h_{50}^{-2}$ erg s$^{-1}$) that is comparable 
to a fossil.

In a study of 6 fossils, Jones et al. (2003) and Jones, Ponman \& Forbes 
(2000) measured the $R$ magnitude difference between the primary elliptical 
galaxy and the second brightest galaxy in the field to be $\Delta$R$_{12}=
2.0-3.3$. They considered all objects within $\sim$ 600 kpc of the 
primary galaxy. Excluding MCG-03-26-030, for the entire observed field around 
each of our isolated 
galaxies, we find $\Delta$R$_{12}=1.9-5.2$ and $\Delta$B$_{12}= 2.3-4.8$.
If we use the maximum common radius for our sample of 116 kpc 
(which corresponds to the closest galaxy NGC 4240), we obtain similar values 
of $\Delta$R$_{12} =1.9-5.3$ and $\Delta$B$_{12}= 2.3-5.4$. 
The galaxy MCG-03-26-030 is the least isolated in our sample with a magnitude 
difference of 1.8 in the $B$-band.
Thus our isolated galaxies have a similar, and some times even greater, 
luminosity isolation than `fossils'.

\subsection{Possible formation scenarios} 

We now discuss the implications of our results in terms of various 
'straw-man scenarios' for how isolated early-type galaxies may have formed. 
Our isolated galaxy images cover hundreds of square kiloparsecs of their 
surrounding fields. Using these images we are not only able to 
study the morphological structure of the galaxies themselves but also the 
magnitude and spatial distribution of the dwarf galaxies in their environments.
Our selection criteria ensure that the galaxies are not currently undergoing 
a strong interaction with any other galaxy.
The possible scenarios for isolated galaxies formation are: 

1) {\it Clumpy collapse at an early epoch}. The early-type galaxy
may have formed at a early epoch from the clumpy collapse or
gaseous merger of many fragments. However the incidence of dust
lanes, plumes and shells indicates that if the galaxy is very
old, then it must have experienced a recent merger/accretion 
event. The low-density environment sample of Kuntschner et
al. (2002) revealed much fine structure and young central ages
indicating recent star formation -- presumably induced by a recent
merger/accretion. Thus if isolated galaxies formed a long time
ago, some of them have undergone strong subsequent evolution.

2) {\it An equal-mass merger of two massive galaxies}. If we consider 
merging of two galaxies as the simple addition of their 
contents, without taking into account any star formation 
during the  merger, then the merger of two 
progenitors of M$_B \sim -20.25$ would produce an elliptical galaxy with 
magnitude of M$_B \sim -21$. The latter is a typical magnitude for our 
isolated galaxy sample. The merger of such
galaxies would be expected to lie on the Kormendy (1977) relation
(Evstigneeva et al. 2004) as we have seen in Fig. 12. 
Furthermore the models of 
Bower et al. (1998) indicate that the merger of equal-mass
galaxies will reproduce the slope and scatter of the CMR as we
observe in Fig. 11. The 44\% of isolated galaxies with dust
would seem to require that at least one of the progenitors was a
spiral galaxy. We note that in a detailed study of NGC 2865, Hau 
et al. (1999) concluded that it is the result of a merger 
involving one spiral and
one early-type galaxy. The plume seen in NGC 1045 also requires
the merger to have involved a spiral galaxy (e.g. Toomre \&
Toomre 1972). In the collisionless 
N-body simulations of Naab \& Burkert (2003), mergers of 
two disks of equal-mass may produce a boxy elliptical, while higher 
mass-ratios may create low-luminosity, rapidly-rotating disky
ellipticals. We find 11\% (one galaxy; NGC 1045) to have boxy 
isophotes and 22\% to have disky isophotes
in our isolated galaxy sample. We conclude that the merger of two
near-equal-mass galaxies, with at least one being a spiral, is a plausible
scenario for several of our isolated galaxies.
 
3) {\it A large elliptical accretes several dwarf
galaxies.} If the elliptical galaxy was initially located on the CMR,  
then the accretion of small dwarfs will not have a 
strong effect on its mass nor on its overall colour. 
A similar argument may apply to the
Kormendy (1977) relation.
However, we would expect low-luminosity dwarfs (with long dynamical
friction timescales) to avoid accretion and be located close to
the isolated galaxy. Such a trend is seen in the  
high central space density for three isolated galaxies
(Fig. 15). Perhaps the main difficulty with this scenario is
explaining the observed fine structure. In three galaxies, the
morphological disturbance (e.g.\ plume, shells, dust lanes) is 
quite strong suggesting the accretion of a fairly large galaxy. 
For the remaining 6 isolated galaxies, this scenario can not be excluded.

4) {\it A group of galaxies collapses with its luminous galaxies
merging together}. Such a scenario would give rise to a 
giant elliptical galaxy (Barnes 1989) 
with a magnitude comparable to the 
total magnitude of the original group. 
The magnitudes of our isolated galaxies
are similar to poor groups but not typical of loose groups which
have a total M$_B \approx -22$. The merged galaxies could explain the
incidence of dust and other fine structure in our sample. However, we would
again expect low-luminosity dwarfs to avoid the merger process, leaving
a giant galaxy surrounded by a population of such dwarfs.
In the case of a virialised X-ray group, its collapse is expected to 
produce an isolated early-type galaxy with an extended group-like X-ray 
emission (called a fossil) because the hot gas cooling time is very long. 
For our sample only NGC 1132, out of four galaxies with published X-ray 
luminosities, has a group-like X-ray emission. Thus most isolated early-type 
galaxies do not appear to be the result of a merged large group, 
but the possibility remains that they may be the product of 
a merged poor group.

\section{Conclusion}  

In this paper we define a new sample of 36 early-type 
isolated galaxies. These galaxies are required to be at least 2 $B$-band 
magnitudes brighter than the 
next brightest galaxy within 0.67 Mpc in the plane of the 
sky and  recession velocity of 700 km s$^{-1}$.
Here we present wide-field imaging for 10 
galaxies of this sample. We also include imaging of four group 
galaxies and an isolated galaxy pair for comparison. 
Our images cover a field-of-view of about 
$30.6 \times 30.6$ arcmin, corresponding to an area of hundreds of square 
kiloparsecs surrounding each galaxy.
From our CCD images we confirm the 
early-type morphology of our sample galaxies.

We used SExtractor to detect galaxies in the fields of the isolated galaxies. 
Assuming the galaxies 
in the field are located at the same redshift as the isolated galaxy, we 
estimate their photometric masses $M$ and distances $R$ from the isolated 
galaxy. 
Excluding all neighbouring galaxies with dynamical times less than 
a Hubble time, we calculated the quantity $M/R^2$ which indicates
the gravitational interaction force applied on the isolated galaxy by 
each galaxy in its local environment. Unlike the galaxies in groups, 
isolated galaxies are relatively gravitationally isolated with no 
neighbours having $M/R^2\gtrsim 4 \times 10^6~ M_\odot/$kpc$^2$.

A mean effective colour of ($B$--$R$)$_e=1.54\pm 0.14$ is measured for our
isolated galaxies which is similar to the colour of early-type galaxies in
denser environments. Also, they are consistent with the slope and smaller
scatter of the colour-magnitude 
relation for Coma cluster galaxies.
This result suggests an early formation epoch for the bulk of stars in our 
isolated and group galaxies.
However, there are also some morphological fine structures 
indicating recent merger/accretion events.
The radial profile of the isophotal parameters, as well as the
model-subtracted images reveal some fine structure such as plumes, 
dust, shells, disky and boxy structures in several isolated galaxies. 
About 44\% of our isolated galaxies show such evidence of a past merger.
The distribution of these different morphological features in our 
isolated galaxies is similar to that of luminous
early-type galaxies in loose groups and clusters, suggesting a similar 
frequency of mergers and interactions.

Our isolated galaxies show a relation between the effective 
radius and the mean surface brightness within that radius 
(the Kormendy relation) which is consistent
with that of luminous ellipticals in groups and clusters. This relation 
constrains the mass ratios of any merger progenitors to be near equal mass.

The galaxies in the fields of the isolated galaxies and 
the groups reveal similar luminosity functions with faint end 
slopes of $\sim -1.2$ and a change in the slope at 
M$_R \approx -18$. This is similar to the luminosity functions obtained by 
other studies in a wide variety of environments.

We also briefly discuss possible formation scenarios for isolated galaxies. 
We conclude that the major merger of two massive galaxies 
could explain several of our sample galaxies.
An alternative scenario of a collapsed poor group of a few galaxies is also 
possible, specifically to explain the observed high central density of dwarfs 
in the field of the isolated galaxy. The collapse of a large, virialized 
group does not appear to explain the most of our isolated galaxies.
Mergers of many gaseous fragments or
the accretion of several dwarfs by a large elliptical galaxy would not give 
rise to the observed fine structures seen in some of our isolated galaxies.

Future spectroscopic observations of the galaxies in the field of 
the isolated galaxies are needed to confirm their redshift, and hence whether 
they are associated with the isolated galaxies or not. High 
signal-to-noise spectra of the isolated galaxy itself will provide
stellar populations and internal kinematics. This will  be very useful to 
obtain a better idea about their star formation histories, and allow us to 
place additional constraints on their possible formation.
\vspace{5mm}

We wish to thank Dr. R. Proctor for his useful comments in the final 
revision of this paper. We are grateful to the supporting team at the Anglo 
Australian Observatory for their help during the observations.
F.M.R acknowledges the 2-year scholarship funding from The Ministry of 
High Education in Egypt. F.M.R. also dedicates this work to the memory of her 
late teacher Prof. A.I. Gamal El Din (NRIAG, Egypt). 
We have used data from the Lyon-Meudon Extragalactic Database (LEDA) compiled 
by the LEDA team at the CRAL-Observatoire de Lyon (France).
This research has made use of the NASA/IPAC Extragalactic Database, which is 
operated by the Jet Propulsion Laboratory, California Institute of Technology, 
under contract with the National Aeronautics and Space Administration.

\section{\bf REFERENCES} \normalsize 
Aars C.E., Marcum P.M., Fanelli M.N., 2001, AJ, 122, 2923\\
Adams M., Jensen E., Stocke J.T., 1980, AJ, 85, 1010\\
Arp H.C., Madore B.F., 1987, A Catalogue of Southern Peculiar Galaxies 
     and Associations. Cambridge Univ. Press, Cambridge\\
Amendola L., DiNella H., Montuori M., Sylos Labini F., 1997, Fractals, 5, 635\\
Balcells, M., 1997, ApJ, 486, 87\\
Balogh M.L., Morris S.L., 2000, MNRAS, 318, 703\\
Barnes J.E., 1989, Nat, 338, 123\\
Barnes J.E., Hernquist L., 1992, ARA\&A, 30, 705\\
Baum W.A., 1959, PASP, 71, 106\\
Bender R., D\"oebereiner S., M\"{o}ellenhoff C., 1988,  A\&AS, 74, 385\\
Bender R., M\"oellenhoff C., 1987, A\&AS, 177, 71\\
Bender R., Surma P., D\"oebereiner S., M\"oellenhoff C., Madejsky R., 1989,
     A\&AS, 217, 35\\
Bernardi M. et al., 2003, AJ, 125, 1882\\
Bertin E., Arnouts S., 1996, A\&AS, 117, 393\\
Binney J., Petrou M., 1985, MNRAS, 214, 449\\
Binney J., Tremaine S., 1987, Galactic Dynamics. Princeton Univ. 
     Press, Princeton, NJ\\
Blanton M.R. et al., 2001, AJ, 121, 2358\\
Blanton M.R. et al., 2003, ApJ, 592, 819\\
Bower R.G., Kodama T., Terlevich A., 1998, MNRAS, 299, 1193\\
Bower R.G., Lucey J.R., Ellis R.S., 1992, MNRAS, 254, 601\\
Bradt H.V., Burke B.F., Canizares C.R., Greenfield P.E., Kelley R.L.,
    McClintock J.E., Koski A.T., van Paradijs J., 1978, ApJ, 226, 111\\	
Capaccioli M., Caon N., D'Onofrio M., 1992, MNRAS, 259, 323\\
Christlein D., Zabludoff A., 2003, ApJ, 591, 764\\
Colbert J.W., Mulchaey J.S., Zabludoff A.I., 2001, AJ, 121, 808 (CMZ01)\\
de Carvalho R.R., Djorgovski S, 1992, ApJ, 389, L49\\
de Vaucouleurs G., de Vaucouleurs A., Corwin H.G., Buta R.J., Paturel G., 
     Fouqué P., 1991, Third reference catalogue of bright galaxies. 
     Springer, New York\\
Dressler A., 1980, ApJ, 236, 351\\
Drory N., Bender R., Feulner G., Hopp U., Maraston C., Snigula J., Hill G.J., 
     2003, ApJ, 595, 698\\
Ellis R.S., Smail I., Dressler A., Couch W.J., Oemler A., Butcher H., 
    Sharples, R.M., 1997, ApJ, 483, 582\\
Evstigneeva E.A., de Carvalho R.R., Ribeiro A.L., Capelato H.V., 2004, 
    MNRAS, 349, 1052\\
Flint K., 2001, PhD Thesis, UC Santa Cruz\\
Flint K., Bolte M., de Oliveira M., 2003, Ap\&SS, 285, 191\\
Forbes D.A., 1991, MNRAS, 249, 779\\
Forbes D.A., Thomson R.C., 1992, MNRAS, 254, 723\\
Franx M., Illingworth G., Heckman T., 1989, AJ, 98,  538\\
Fujita Y., 2004, PASJ, 56, 29\\
Garcia A.M., 1993, A\&AS, 100, 47\\
Girardi M., Mardirossian F., Marinoni C., Mezzetti M., Rigoni E., 2003, 
A\&A, 410, 461\\
Gladders M.D., Lopez-Cruz O., Yee H.K.C., Kodama T., 1998, ApJ, 501, 571G\\
Graham A., Colless M., 1997, MNRAS, 287, 221\\
Goudfrooij P., de Jong T.,  Hansen L., N{\o}rgaard-Nielsen H.U., 1994a,
    MNRAS, 271, 833\\
Goudfrooij P., Hansen L., J{\o}rgensen H.E., N{\o}rgaard-Nielsen H.U., de
    Jong T., van den Hoek L.B., 1994b, A\&AS, 104, 179\\
Gunn J.E., Gott J.R., 1972, ApJ, 176, 1\\
Hamabe M., Kormendy J., 1987, in de Zeeuw T., ed., Proc. IAU Symp. 127, 
Structure and Dynamics of Elliptical Galaxies. Reidel, Dordrecht, p. 379\\
Hau G.K.T., Carter D., Balcells M., 1999, MNRAS, 306, 437\\
Hoessel J.G., Oegerle W.R., Schneider D.P., 1987, AJ, 94, 1111\\
Horner D., Goudfrooij P., 2004, submitted to AJ \\
Jedrzejewski R.I., 1987, MNRAS, 226, 747\\ 
Jones L.R., Ponman T.J., Forbes D.A., 2000, MNRAS, 312, 139. \\
Jones L.R., Ponman T.J., Horton A., Babul A., Ebeling H., Burke D.J., 2003, 
MNRAS, 343, 627\\
Karachentseva V.E., 1973, Comm. Spac. Ap. Obs. USSR, 8, 1\\
Khochfar S., Burkert A., 2003, ApJ, 597, 117\\
Khosroshahi H.G., Raychaudhury S., Ponman T.J., Miles T.A., Forbes D.A., 
     2004, MNRAS, 349, 527\\
Kodama T., Bower R.G., Bell E.F., 1999, MNRAS, 306, 561\\
Kodama T., 2001, A\&SS, 276, 877\\
Kormendy J., 1977, ApJ, 218, 333\\
Kuntschner H., Smith R.J., Colless M., Davies R.L., Kaldare R., Vazdekis A., 
2002, MNRAS, 337, 172\\
La Barbera F., Busarello G., Merluzzi P., Massarotti M., Capaccioli M., 
   2003, ApJ, 595, 127\\
Landolt A.U., 1992, AJ, 104, 340\\
Lauberts A., Valentijn E.A., 1989, The surface photometry catalogue of the 
ESO-Uppsala galaxies. European Southern Observatory (ESO), Garching\\
Lauer T.R., 1985, MNRAS, 216, 429\\
Merluzzi P., 1998, A\&A, 338, 807\\
Miles T.A., Raychaudhury S., Forbes D.A., Goudfrooij P., Ponman T.J., 2004, 
MNRAS, submitted\\
Mobasher B., Colless M., Carter D., Poggianti B.M., Bridges T.J., 
      Kranz K., Komiyama Y., Kashikawa N., Yagi M., Okamura S., 
      2003, ApJ, 587, 605\\
Moore B., Katz N., Lake G., Dressler A., Oemler A., 1996, Nat, 379, 613\\
Moore B., Lake G., Katz N., 1998, ApJ, 495, 139\\
Moore B., Lake G., Quinn Th., Stadel J., 1999, MNRAS, 304, 465\\
Mulchaey J.S., Zabludoff A.I., 1999, ApJ, 514, 133\\
Naab T., Burkert A., 2003, ApJ, 597, 893\\
Nieto J.-L., Bender R., 1989, A\&A, 215, 266\\
O'Sullivan E., Forbes D.A., Ponman T.J., 2001, 328, 461\\
Peletier R.F., Davies R.L., Illingworth G.D., Davis L.E., Cawson M., 1990, AJ, 
100, 1091\\
Pfefferkorn F., Boller Th., Rafanelli P., 2001, A\&A, 368, 797\\
Ponman T.J., Allan D.J., Jones L.R., Merrifield M., McHardy I.M., Lehto H.J., 
      Luppino G.A., 1994, Nat, 369, 462\\
Prugniel P., Heraudeau P., 1998, A\&AS, 128, 299\\ 
Prugniel P., Simien F., 1997, A\&A, 321, 111\\
Raychaudhury S., 1989, Nat, 342, 251\\
Raychaudhury S., 1990, BAAS, 22, 1331\\	
Reduzzi L., Longhetti M., Rampazzo R., 1996, MNRAS, 282, 149 (RLR96)\\
Saucedo-Morales J., Bieging J., 2001, ApSSS, 277, 449\\
Schechter P., 1976, ApJ, 203, 297\\
Schlegel D.J., Finkbeiner D.P., Davis M., 1998, ApJ, 500, 525\\
Schweizer F., Seitzer P., Faber S.M., Burstein D., Dalle Ore C.M., Gonzalez 
J.J., 1990, ApJ, 364L, 33\\
Smith R.M., Martinez V.J., Graham M.J., 2003, preprint (astro-ph/0311599) 
(SMG03)\\
Soares D.S.L., de Souza R.E., de Carvalho R.R., Couto de Silva T.C., 1995, 
A\&AS, 110, 371\\
Stanford S.A., Eisenhardt P.R.M., Dickinson M.E., 1995, ApJ, 450, 512\\
Stanford S.A., Eisenhardt P.R.M., Dickinson M.E., 1998, ApJ, 492, 461\\
Stocke J.T, Keeney B.A., Lewis  A.D., Epps H.W., Schild R.E., 2004, AJ, 
127, 1336\\  
Terlevich A.I., Forbes D.A., 2002, MNRAS, 330, 547\\
Toomre A., Toomre J., 1972, ApJ, 178, 623\\
Trentham N., Hodgkin S., 2002, MNRAS, 333, 423\\
Trentham N., Tully R.B., 2002, MNRAS, 335, 712\\
Waddington I., Windhorst R.A., Cohen S.H., Dunlop J.S., Peacock J.A., 
      Jimenez R., McLure R.J., Bunker A.J., Spinrad H., Dey A., Stern D.,
      2002, MNRAS, 336, 1342\\
Zabludoff A., \& Mulchaey J., 2000, ApJ, 539, 136\\
Zepf S.E., Whitmore B.C., 1993, ApJ, 418, 72\\
Ziegler B.L., Saglia R.P., Bender R., Belloni P., Greggio L., Seitz S., 1999, 
A\&A, 346, 13\\    
Zwicky F. et al., 1957, Catalog of Galaxies \& Clusters of Galaxies, Vol. 1-6, 
California Institute of technology, Pasadena\\

\end{document}